\documentclass[fleqn,usenatbib]{mnras}

\usepackage{newtxtext,newtxmath}

\usepackage[T1]{fontenc}

\DeclareRobustCommand{\VAN}[3]{#2}
\let\VANthebibliography\thebibliography
\def\thebibliography{\DeclareRobustCommand{\VAN}[3]{##3}\VANthebibliography}

\usepackage{graphicx}	
\usepackage{amsmath}	

\usepackage{multirow}
\usepackage{diagbox}
\usepackage
{caption}
\usepackage{subcaption}

\usepackage{orcidlink}
\usepackage{xcolor}

\usepackage{hyperref}
\usepackage{appendix}

\usepackage{tabularx}  

\newcommand{\citeg}[1]{\citep[e.g.,][]{#1}}

\newcommand{\aref}[1]{\hyperref[#1]{Appendix~\ref{#1}}}
\defcitealias{vijayan24}{QED I}

\title[X-ray metallicity gradients]{\textsc{Quokka}-based understanding of outflows (QED) - II. X-ray metallicity gradients as a signature of galactic wind metal loading}

\author[R. Huang et al.]{Rongjun Huang\orcidlink{0000-0002-6646-8365}$^{1, 2}$\thanks{E-mail: U6569836@anu.edu.au, Astro@Rongjun-Huang.com}, 
Aditi Vijayan\orcidlink{0000-0002-7714-2379}$^{1, 2}$\thanks{E-mail: Aditi.Vijayan@anu.edu.au}, 
Mark R. Krumholz\orcidlink{0000-0003-3893-854X}$^{1, 2}$\thanks{E-mail: Mark.Krumholz@anu.edu.au} 
\\
$^{1}$Research School of Astronomy and Astrophysics, Australian National University, Cotter Road, Weston Creek, ACT 2611, Australia\\ 
$^{2}$ARC Centre of Excellence for All Sky Astrophysics in 3 Dimensions (ASTRO 3D), Australia\\
}

\date{Accepted XXX. Received YYY; in original form ZZZ}

\pubyear{2025}

\begin{document}
\label{firstpage}
\pagerange{\pageref{firstpage}--\pageref{lastpage}}
\maketitle

\begin{abstract}
Supernova-driven galactic outflows play a vital but still poorly-understood role in galactic chemical evolution, and one of the largest uncertainties about them is the extent to which they consist of supernova ejecta that are unmixed, or only poorly-mixed, with the remainder of the interstellar medium (ISM). Simulations of wind launching make a range of predictions about the extent of mixing between the wind and the ISM, but thus far these have proven challenging to test observationally. In this study, we post-process high-resolution simulations of outflows from the QED simulation suite to generate synthetic X-ray spectra from galactic winds, which we then analyse using standard observational procedures, in order to search for detectable markers of wind mixing. Our synthetic observations reveal that partially-mixed winds show significant and detectable metallicity gradients when viewed edge-on, with metallicity decreasing away from the central galactic disc. We explore how this signature results from imperfect mixing and the extent to which measurements of it can be used to diagnose the level of mixing in winds. We determine the signal-to-noise ratio (SNR) requirements for such measurements to be reliable, and provide a simple quantitative model that can be used to connect metallicity gradients to mixing between the hot ($T>10^{6}$ K) and cold ($T<10^{4}$ K) phases in observations that reach the required SNR, providing a framework to interpret current and future observations. 
\end{abstract}




\section{Introduction}

Metals are synthesized in the cores of stars that, at least in spiral galaxies such as the Milky Way, reside in a thin disc. However, observations reveal that these metals are not confined to the disc. Instead, $\sim 50\%$ of the mass of metals like oxygen is found in galactic halos, at distances as large as $\sim 100\ \text{kpc}$ from the central galaxy \citep{Tumlinson11a, peeples14}. This raises the intriguing question of how metals are transported from the disc to the halo. The most plausible explanation is that they are carried by galactic outflows, powerful winds driven by supernova (SN) feedback that can transport material, momentum and energy from the disc to the halo. 

Physically, these outflows are driven by SN thermal energy injection that heats up the surrounding gas, driving it to expand and break out of the disc. Once hot bubbles break out, the momentum-carrying gas expands freely and eventually ends up in the circum-galactic medium (CGM) around the galaxy. Because this hot gas is generated as a direct result of SN activity, it is expected to be loaded with metals and therefore enriches the CGM with metals. However, the balance in the gas escaping the galaxy between unmixed supernova ejecta and swept-up ambient gas from the galactic interstellar medium (ISM) -- which is also metal-enriched but much less so than the fresh supernova ejecta -- is highly uncertain. It is also highly important: the fraction of SN-produced metals that are promptly lost to the outflow rather than being retained in the disc is a crucial free parameter in models of galactic chemical evolution \citep[e.g.,][]{Peeples11a, Telford19a, Forbes19a, Johnson20a, Sharda21a, Kravtsov22a}.

One potential way to quantify the fraction of SN-injected metals carried by the outflowing gas is by studying the diffuse X-ray emission around star-forming galaxies. Such emission is linked to star-formation processes occurring within the disc \citep{Strickland+04, Grimes+05, Tullmann+06, Owen+09, Li&Wang2013, Wang+16}, and is useful for studying metallicity due to the large number of abundance-diagnostic lines that appear in soft X-rays. In particular, the starburst M82 serves as a textbook case of SN-driven outflows and much attention has been given to measuring the metallicity of the outflows launched from the central region in the X-ray \citeg{Read02a, Origlia04a, ranalli08, konami11, lopez20, fukushima24}. All these studies have revealed that the outflows M82 have non-uniform metallicities. \citet{Read02a} and \citet{Origlia04a} found that the outflows are enhanced in $\alpha/\mathrm{Fe}$ relative to the disc, while \cite{lopez20} measured the gradients of metallicity along the minor axis of M82 and reported a clear trend of declining abundances and gas temperatures with distance away from the central galaxy for several $\alpha$ elements. They hypothesize that the outflow profile of M82 is a result of mixing between different temperature phases and not simply because of the adiabatic expansion of hot wind. However, the nature of X-ray emission poses challenges for interpretation. For example, spectral fitting relies on simplifying assumptions about temperature profiles of X-ray emitting gas \citep{Vijayan&Li}. One way to overcome such limitations is to provide a theoretical basis for the interpretation of metallicity gradients using simulations.

While M82 is the best-studied case, thanks to its proximity and nearly edge-on geometry, there is also substantial evidence for metallicity variations in the winds of other galaxies, and in phases other than the hot, X-ray-emitting one. \citet{Lopez23a} find a metallicity gradient in the hot phase of the wind of NGC 253 that is similar to that in M82. \citet{cameron21} identify a clear negative metallicity gradient along the minor axis of the wind of edge-on starburst galaxy Markarian 1486 using direct (electron temperature-based) measurements of metallicity in the warm ionised phase. \citet{Martin02a} report significant $\alpha$/Fe enhancement in the hot wind phase of the dwarf starburst NGC 1569, but \citet{Hamel-Bravo24a} find no comparable enhancement in the warm ionised gas, suggesting a composition difference between these two components. \citet{Xu22a} also conclude that there is likely a significant difference between the metal content of the warm ionised and hot components of the wind in their absorption spectroscopic study of 45 low-redshifts starbursts drawn from the CLASSY sample \citep{Berg22a}. \citet{chisholm18} carry out absorption spectroscopy in a sample of seven local galaxies and find evidence that metallicities in the warm ionised wind phase are significantly enhanced compared to the interstellar media of the galaxies driving those winds. Significantly, their sample contains ordinary star-forming galaxies in addition to starbursts, and there is no significant difference in the metal enhancement between the two groups. This suggests that, while most studies of metal-enriched winds to date have focused on starbursts simply because their winds are bright and easy to observe, metal-enrichment is a phenomenon common to all galactic winds, not just the winds of starbursts.

There have also been numerous attempts to use simulations to study the metal content of galactic winds. This problem is generally impossible to address in cosmological simulations, which lack the resolution to capture the phase structure of winds without relying on sub-grid models \citep[e.g.,][]{Smith24a}. However, even excluding cosmological simulations, there are a range of approaches that represent different trade-offs between volume, resolution, and physical complexity. At the finest resolution but lowest volume end are ``tall-box'' simulations \citep[e.g.,][]{kim20a, Rathjen21a} that simulate only a portion of a galactic disc, while at the larger-volume, lower-resolution end are simulations of entire isolated dwarf galaxies \citep[e.g.,][]{melioli13, emerick18, Emerick19a, Emerick20a, Schneider18b, schneider20, andersson23, rey24, steinwandel24, Steinwandel24a}. Similarly, there are a range of levels of physical complexity, with some simulations using relatively simple treatments of ISM physics in order to maximise resolution \citep[e.g.,][]{melioli13, Schneider18b, schneider20} while others include live treatments of physical processes such as radiative transfer \citep[e.g.,][]{rey24} and cosmic rays \citep[e.g.,][]{Rathjen21a} or including multiple elements with different nucleosynthetic origins \citep[e.g.,][]{emerick18, Emerick19a, Emerick20a}. Moreover, recent high-resolution dwarf galaxy simulations have incorporated multiple feedback channels \citep[e.g.,][]{smith21, gutcke21, deng24} or focused on star-by-star physics in isolated systems \citep[e.g.,][]{hu16, hu17, hu19, hu23a, hu23b}, highlighting the importance of resolving supernova remnants and properly modeling radiation in shaping multiphase, metal-enriched outflows. However, there have thus far been limited efforts to carry out detailed observational comparisons between these simulations and X-ray observations; the primary exception is the work of \citet{schneider24}, who generate detailed synthetic X-ray maps from their simulations. However, those simulations do not include metals, and thus are not directly usable for comparing to observations of metallicity variation in winds such as those summarised above.

Part of the reason for this relative paucity of studies of the metal content of winds including detailed observational comparisons is the numerical difficulty of the problem. Simulations targeting these observations require very high resolution because mixing between the hot gas that emits in X-rays and the cooler gas that can be studied in optical or radio is a critical process in determining both the temperature and the metal content of the hot phase. The rate of mixing between the hot and cooler phases is easily affected by poor resolution, with Lagrangian methods tending to under-mix without an explicit sub-grid diffusion model \citep{Shen10a, Escala18a}, and Eulerian methods tending to over-mix due to numerical diffusion. At a minimum, no simulation that has to rely on sub-grid models of SN explosions to avoid over-cooling can reliably study metal mixing, a condition that rules out even zoom-in cosmological simulations, and even properly resolving the Sedov-Taylor phase may be insufficient to yield converged results for clustered SNe \citep{Gentry19a}. At the same time, observations of wind metal content generally target regions extending to several kpc off galactic discs, and thus simulations attempting to compare to these observations must have similar extents. Few existing simulations meet these simultaneous requirements. 

In this work, we use QED\footnote{\textsc{QUOKKA}-based Understanding of Outflows Derived from Extensive, Repeated, Accurate, Thorough, Demanding, Expensive, Memory-consuming, Ongoing Numerical Simulations of Transport, Removal, Accretion, Nucleosynthesis, Deposition, and Uplifting of Metals (QUOD ERAT DEMONSTRANDUM, or QED)} simulations, introduced in \citet[hereafter \citetalias{vijayan24}]{vijayan24}, to study metals in galactic winds with mock X-ray observations. QED is a suite of 3D hydrodynamic simulations of galactic winds generated by the new code \textsc{Quokka} \citep{wibking22, He24a}, a state-of-the-art adaptive mesh refinement (AMR) radiation-hydrodynamics (RHD) tool optimized for graphics processing units (GPUs). In terms of the taxonomy of simulation approaches discussed above, the QED suite lies at the maximum resolution end of the spectrum: tall box rather than whole-galaxy, and including a relatively simple treatment of ISM physics and feedback that includes only driving by randomly-distributed SNe. QED's primary virtue is the very large dynamic range afforded by GPU acceleration: the simulations feature a uniformly high resolution of $2$ pc that allows us to accurately follow the metal exchanges taking place in the multi-phase outflows. \citetalias{vijayan24} demonstrates that, at the 2 pc resolution achieved in the QED suite, the wind mass flux, metal flux, and metallicity as a function of height off the midplane are all numerically converged. At the same time, the simulation domain extends to $\pm 4$ kpc above and below the galactic midplane, enabling us to follow trends up to a few kpc above the disc, capturing the off-disc regions typically targeted in observations. Of course QED -- like all tall box simulations -- still captures only a small fraction of the volume of a real galactic wind in a galaxy like the Milky Way, which has an area much larger than 1 kpc$^2$ area of the QED tall box. However, it is reasonable to hypothesize that the QED simulation box is statistically representative of the full wind, and is therefore suitable for use in simulated observations provided that we compensate for the smaller volume. Finally, the \textsc{Quokka} code on which the QED simulations are based uses a very high-resolution hydrodynamics scheme based on a PPM solver with a monotonized-central (MC) limiter, using a combination of a dual energy formalism and first-order flux correction to maintain stability in problematic cells \citep{wibking22}; this scheme should be significantly less dissipative, and thus better able to follow phase and metal mixing, than the more dissipative methods used in some earlier simulations.

This paper is organized as follows:
\autoref{sec:methods} provides a brief summary of the setup of the QED simulations, and the method we use to generate synthetic observations from them and then analyse them using the same techniques that are routinely applied to real observations. \autoref{Results and Analysis} presents our main results related to fits of the mock observations, and \autoref{Discussion} discusses the interpretation of the results and their implications. We outline our major conclusions in \autoref{sec:conclusion}.

\section{Methods and Data}\label{sec:methods}

In this section, we first introduce the QED simulations (\autoref{sec:qed_data}) and then explain our methods for generating synthetic X-ray data (\autoref{sec:xray-data}) and analysing those synthetic data to extract physical parameters (\autoref{sec:fitting_xray_spectrum}).

\subsection{The QED simulations}\label{sec:qed_data}
\begin{figure*}
     \centering
    $$
    \begin{array}{c}
        \includegraphics[width=2\columnwidth]{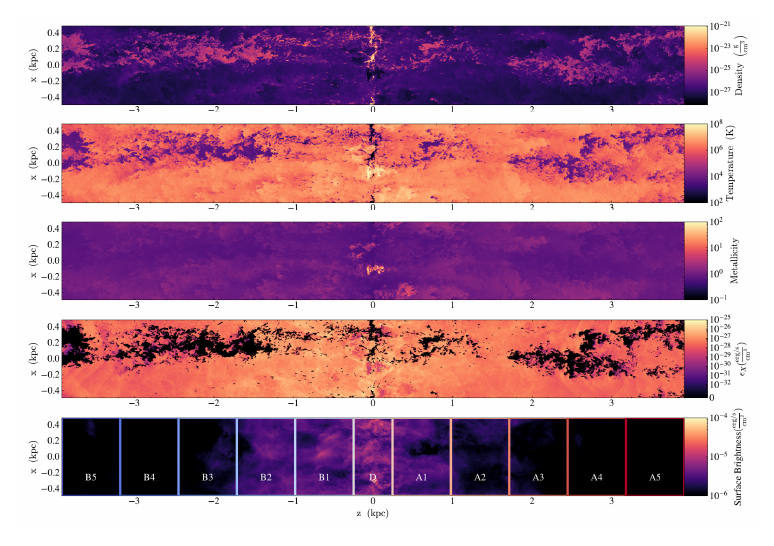}
    \end{array}
    $$
    \caption{Slices in the $xz$ plane from a snapshot of the QED simulations \citepalias{vijayan24} at $110$ Myr. From the top to bottom we show gas density, temperature, metallicity, and X-ray emissivity field per unit volume in the $0.4 - 2$ keV band in the first four panels. The bottom panel shows the X-ray surface brightness, overlaid with rectangles highlighting the 11 regions over which we average to produce the spectra used for analysis. The metallicity and X-ray luminosity panels are computed using $Z_\mathrm{bg} = 0.5Z_\odot$. }     
    \label{fig:5plots}
\end{figure*}

QED is a suite of fixed-grid simulations of a $\sim 1\times 1$ kpc$^2$ patch of the Milky Way disc; \citetalias{vijayan24} presents results and a convergence study for Solar neighbourhood conditions, while \citet{vijayan_prep} extends the model grid to a wide range of galactic environments. In this paper, we use the highest-resolution case presented in \citetalias{vijayan24}, and we refer readers to that paper for full details of the numerical setup, which we only summarise here. The simulation has a simulation box centered on the disc midplane that extends to $\pm 4$ kpc and a uniform resolution of $512 \times 512 \times 4096$, giving a grid spacing of $2$ pc, and is run for 116 Myr, by which point the outflow has reached statistical steady state. We show mass outflow rates and loading factors as a function of time and simulation resolution to demonstrate convergence in \autoref{app:convergence}. The gas is confined by the gravity of the stellar disc and dark matter halo and driven by SNe, which occur at a constant rate calibrated to the \citet{kennicutt98} relation and distributed exponentially in height around the disc with a scale height of 150 pc. SNe are treated as injections of $10^{51}$ erg of pure thermal energy; the resolution is high enough that the Sedov-Taylor phase is well-resolved, and no sub-grid models are required. Gas cooling is implemented through a tabulated cooling function derived from the \textsc{Grackle} cooling model \citep{smith17}. After nearly $100$ Myr of SN feedback, the simulation reaches quasi-steady-state outflow rates. We show $y=0$ density and temperature slices from a snapshot in $110$ Myr in the top two panels of \autoref{fig:5plots}. The outflows clearly possess a multi-phase structure characterised by a broad range of temperatures from $<10$ K to $>10^7$ K, with high (low) temperature gas being associated with low (high) densities. 

Metals in the simulations can be either injected by SNe or present in the initial conditions. For the former, each SN injects a fixed mass $\Delta M_\mathrm{SN} = 1$ M$_\odot$ of oxygen into the same cell in which it deposits thermal energy, which is subsequently advected as a passive scalar. The second source of metals is the initial background metallicity of the ISM present at the start of the simulation, which we parameterise by $Z_{\rm bg}$. We scale it to Solar abundances, so the initial oxygen mass density in a cell of total mass density $\rho$ is $\rho_\mathrm{O} = \rho (Z_\mathrm{bg}/Z_\odot) Z_\mathrm{O,\odot}$, where $Z_\odot$ denotes Solar metallicity and we set the Solar oxygen abundance to $Z_\mathrm{O,\odot} = 8.6\times 10^{-3}$. We treat $Z_{\rm bg}$ as an independent parameter that we can alter in the post-processing of the simulations.\footnote{We note here that in \citetalias{vijayan24} all gas cools at Solar metallicity, irrespective of the local metallicity. This is the downside to the flexibility we gain by our purely passive treatment of metals.} The total metal density of a cell is the sum of the background and SN-added contributions -- see Section 2.3.1 of \citetalias{vijayan24} for full details of the computation. Once we have computed the oxygen metallicity in a cell, we assume Solar-scaled abundances for all other elements. This assumption is not entirely realistic, since we are modeling only type II SNe that produce predominantly $\alpha$ elements rather than type Ia SN that produce iron peak elements, and thus our added metal field should have a different $\alpha$/Fe ratio than our background field; we discuss the implications of this in more detail below. 

In the third panel from the top of \autoref{fig:5plots}, we show the metallicity of the same slice whose density and temperature are shown in the top two panels. We note here that higher metallicity is correlated with higher temperatures and lower densities tracing the SN-processed gas. Lower metallicity mainly follows the high-density warm gas. 

\subsection{Generation of synthetic X-ray data}
\label{sec:xray-data}

We use the X-ray post-processing packages \texttt{pyXSIM} and \texttt{SOXS} \citep{zuhone16, zuhone23} based on the \texttt{yt} analysis and visualisation suite \citep{turk11} to generate synthetic observations from a simulation snapshot via a procedure we describe in this Section.

\subsubsection{Generating photons using \texttt{pyXSIM}}\label{subsec:pyxsim}

The \texttt{pyXSIM} package generates a set of X-ray photons from each cell in the simulation under the assumption of a collisionally ionized and optically thin plasma, following the Astrophysical Plasma Emission Code \citep[APEC;][]{smith01}. Because we are interested in the diffuse X-ray emission from galactic winds, we limit the photon energies to the soft X-ray band, with energies $0.4 - 2.0$ keV; this is the energy range over which \textit{Chandra} is the most sensitive, contains most (but not all) of the important diagnostic lines for metallicity, and in practice is usually the only usable band for spatial studies of galactic winds because emission from the relatively harder X-ray band ($2-5$ keV) is significant compared to the soft X-ray contribution only in regions close to the disc, and drops sharply away from it \citep{Vijayan+18}. One complication in generating synthetic observations from our simulation is that we simulate only a 1 kpc$^2$ patch that represents a small fraction of an actual Milky Way-sized galaxy (typical area closer to 100 kpc$^2$).\footnote{Conversely, our simulation volume would be more realistic for a circumnuclear starburst such as M82. However, in such a galaxy the wind would be much denser and more powerful than in our Solar neighbourhood setup, and thus much brighter in the X-rays.} In order to compensate for this, we tile 100 copies of our simulation box together in order to generate our synthetic spectra -- that is, we generate 100 times as many X-ray photons from each cell per unit time as would be expected, on the basis that a realistically-large galaxy would contain 100 times as many cells with similar physical conditions.

In addition to gas density and temperature, \texttt{pyXSIM} also requires metallicity. As discussed in \autoref{sec:qed_data}, this depends on the choice of initial background metallicity $Z_\mathrm{bg}$; we carry out our analysis for the cases $Z_\mathrm{bg} = 0$, 0.2, 0.5, 1, and 2, providing a wide baseline. We summarise our choices for the other parameters required by \texttt{pyXSIM}, including the exposure time, collecting area, and distance to source, in \autoref{table:pyXSIM_parameters}. Our choice of collecting area is typical of \textit{Chandra}, and our distance is roughly the distance to M82, and is similar to the distances of other nearby galaxies whose wind metallicities have been measured in X-rays (e.g., NGC 1569 -- \citealt{Martin02a}; NGC 253 -- \citealt{Lopez23a}). In order to get a ``pure'' unadulterated spectrum, we disable absorption, backgrounds, and foregrounds, and we use a large exposure time of 10 Ms. We will consider the effects of smaller and more realistic signal-to-noise ratios (SNRs) in \autoref{sec:impact_of_snr}. We do not seek to explore the effects of confusion due to absorption, backgrounds, and foregrounds because these are all highly variable across the sky, and thus it is not possible to make general statements about the extent to which they can be corrected and the level of observational uncertainty introduced by the need to make such corrections. 


\autoref{fig:5plots} shows the $0.4 - 2.0$ keV X-ray luminosity per unit volume ($\epsilon_X$, second from the bottom panel) and the surface brightness (bottom panel) as computed by \texttt{pyXSIM} for the same 110 Myr snapshot shown in the upper panels, and for $Z_\mathrm{bg} = 0.5Z_\odot$. As expected, the X-ray luminosity is spatially correlated with gas at temperatures $\gtrsim 10^5$ K, and for gas at these temperatures is positively correlated with density. 

\subsubsection{Characteristics of the X-ray emitting gas}

Before proceeding to simulated observations, it is important to examine the characteristics of the X-ray emitting gas, both to ensure that they are physically reasonable and to ensure that we adequately resolve the emitting regions in our simulations. The reason that resolution is a potential concern is that, thanks to the strong density and temperature dependence of X-ray emissivity, in a multiphase medium such as our simulated wind, X-ray emission can be dominated by relatively thin interface layers between the hot volume-filling phase and the cooler entrained gas that can be hard to resolve \citep{toala18, Vijayan+18}.

To this end, we examine the cumulative distribution functions (CDFs) of X-ray emission with respect to gas temperature $T$ and volume $V$ for the same snapshot (and using the same $Z_\mathrm{bg}=0.5Z_\odot$ as shown in \autoref{fig:5plots}. For the purpose of analysis discussed here and in subsequent Sections, we partition the simulation domain into 11 regions, mirroring the approach used by \cite{lopez20}; we mark our regions in the bottom panel of \autoref{fig:5plots}. The central region, labeled `D', corresponds to the area where SNe are injected and extends up to a height of 489.03 pc (i.e., 256 cells), while the remaining 10 regions (`A/B1' to `A/B5') cover the outflow zone, each spanning 733.55 pc (i.e., 384 cells) in height.

\begin{figure*}
     \centering
    $$
    \begin{array}{cc}
    \centering
    \includegraphics[width=\columnwidth]{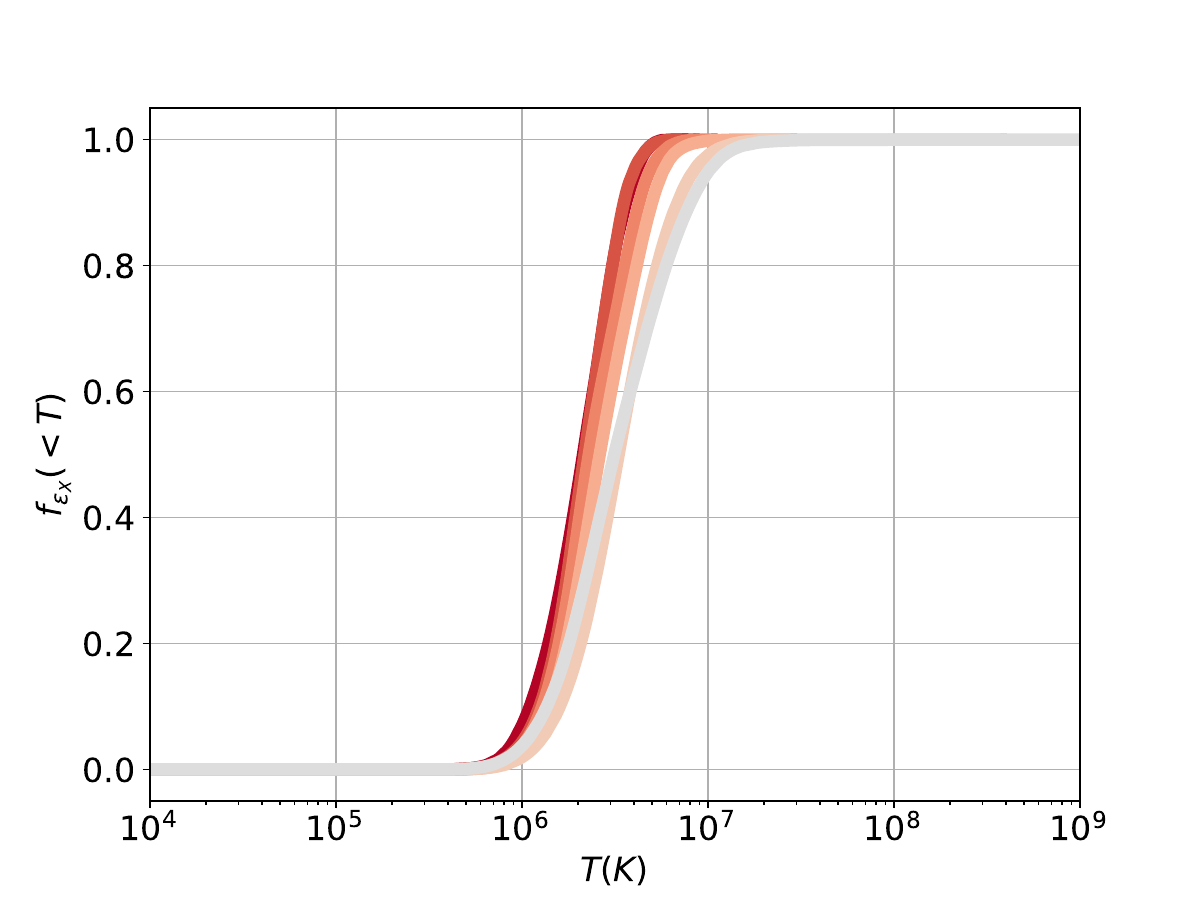} & 
    \includegraphics[width=\columnwidth]{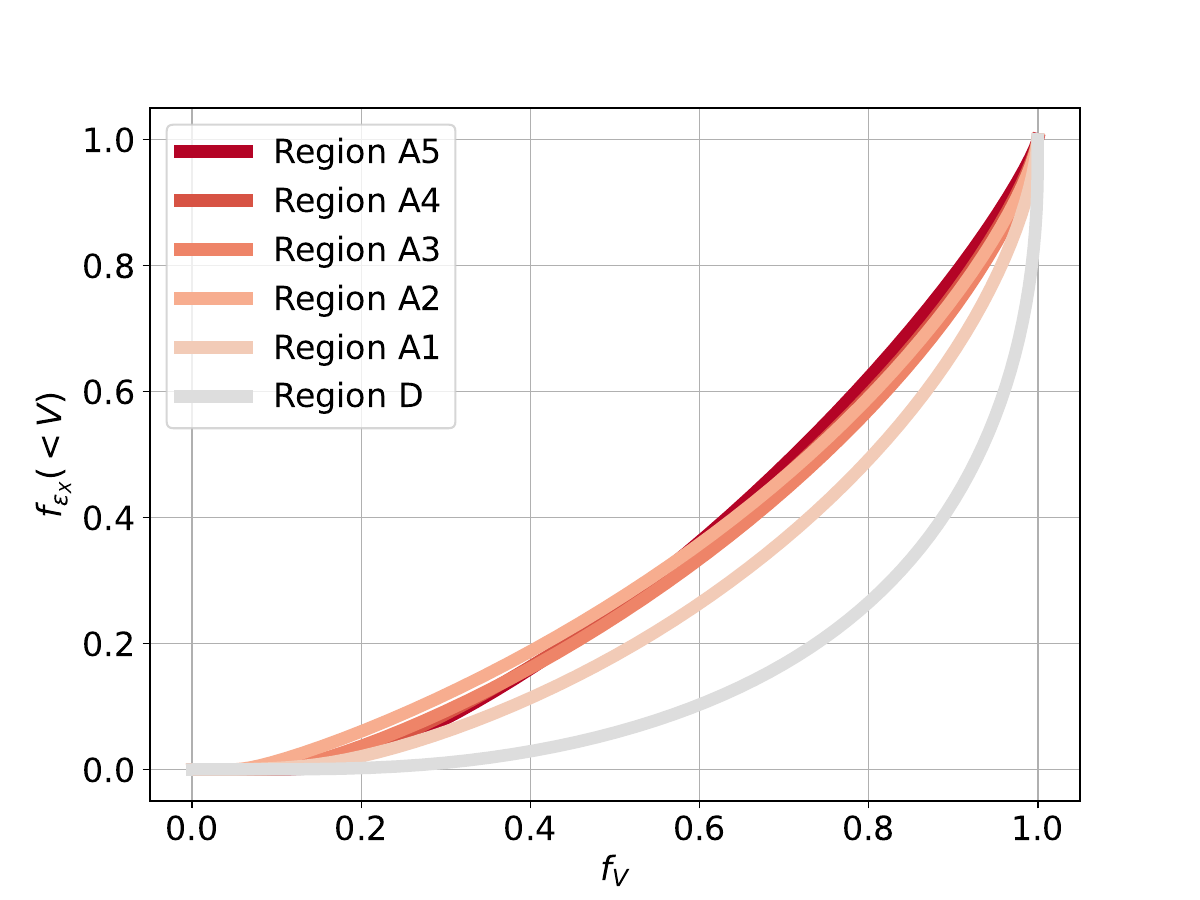}
    \end{array}
    $$
    \caption{Cumulative distribution of X-ray emission in the 0.4–2.0 keV band with respect to gas temperature (left) and volume (right). Different colours correspond to different regions within the simulation, as explained in the main text; we show only the `A' regions at positive $z$ and omit the `B' regions at negative $z$ to reduce clutter in the plot, but the results for the `A' and `B' regions are qualitatively identical. We see that the X-ray emitting gas lies within a narrow temperature range ($\approx 10^6 - 10^7$ K), but that the emitting regions are reasonably well-resolved, with $50\%$ of the emission coming from $\approx 30\%$ of the volume in all regions but D, where the emission arises from $\approx 10\%$ of the volume.}
    \label{fig:frac_epsilon_vs_frac_V_&_T}
\end{figure*}

We show the CDFs we obtain for each region in \autoref{fig:frac_epsilon_vs_frac_V_&_T}. The two panels of this Figure show the fraction of total X-ray emission in the 0.4-2.0 keV band arising from gas with temperature $<T$ (left panel) and from a a given fraction of total volume (right). As expected based on the models of \citet{toala18}, the left panel of this Figure shows that X-ray emission in our simulation is dominated by moderately ($\sim$ few $10^6$ K) rather than extremely ($\gtrsim 10^7$ K) hot gas.
The right panel, however, shows that the volume occupied by this gas is substantial, and is thus well-resolved within our simulations.
Quantitatively, except in region `D', we find that $\approx 50\%$ of the emission emerges from 30\% of the volume. Given that our simulation contains a total of $\approx 10^9$ computational cells, even 30\% of this number constitutes very good resolution: quantitatively, 30\% of the volume corresponds to a resolution of roughly $685^3$ cells. Resolution is slightly worse in region `D', where 50\% of the emission emerges from just under 10\% of the volume, but even 10\% of our simulation volume corresponds to $475^3$ cells. The lower volume filling fraction in region `D' is not surprising since this region is dominated by fresh SNe ejecta with very high metallicities. We therefore conclude that the emitting regions in our simulations are relatively well-resolved. 

\begin{figure*}
     \centering
    $$
    \begin{array}{c}
        \centering
        \includegraphics[width=2\columnwidth]{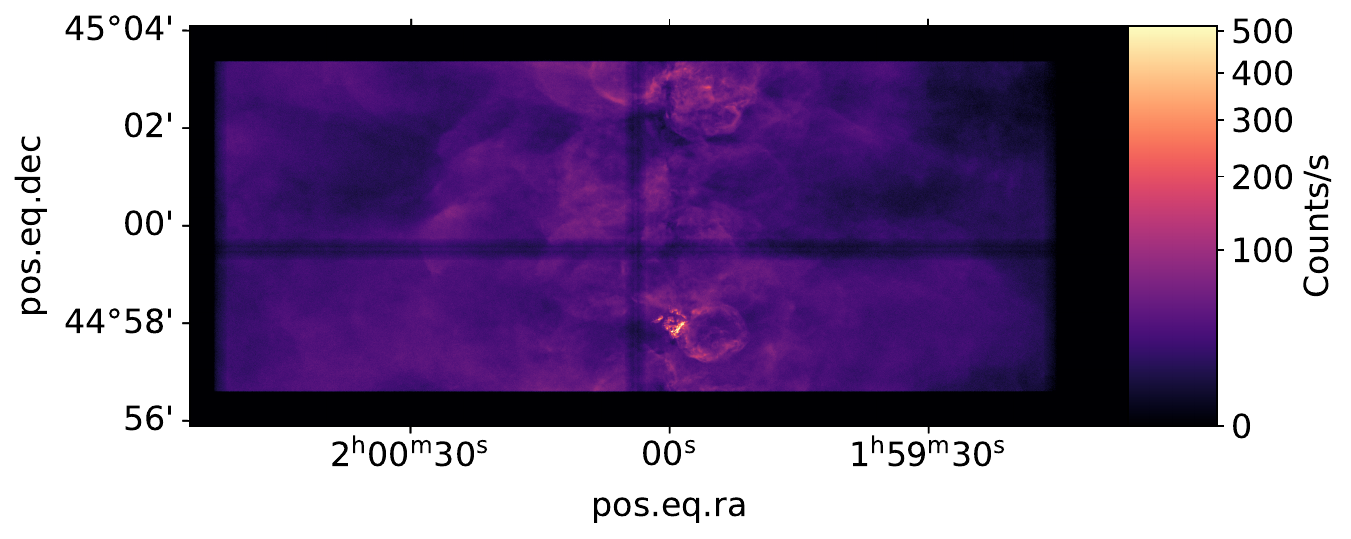} 
    \end{array}
    $$
    \caption{A simulated \textit{Chandra}/ACIS-I image in the 0.4 - 2 keV band of the snapshot shown in \autoref{fig:5plots}; the colorbar shows photon count rate per pixel. We take $Z_\mathrm{bg} = 0.5Z_\odot$ and use a 10 Ms exposure time.}       
    \label{fig:soxs_hot}
\end{figure*}

\subsubsection{Generating mock observations using \texttt{SOXS}}

Once the X-ray photons are generated, we use \texttt{SOXS} to produce mock images and spectra as they would be observed by the \textit{Chandra X-ray Observatory}. \texttt{SOXS} uses a subset of these photons to project on the plane of the instrument and convolve with the instrument response to produce an events file. We use the response matrix (specified by Auxiliary Response File and Redistribution Matrix File) of the ACIS-I instrument which has a FoV of $16.9\times16.9$ arcmin$^2$ and a spectral resolution of 130 eV at 1.49 keV. The mock \texttt{SOXS} observation also needs an exposure time; we carry out our post-processing for exposure times of 10, 1, 0.1, and 0.01 Ms in order to generate results with a range of SNRs. We summarise the parameters we use for the \texttt{SOXS} post-processing in \autoref{table:SOXS_parameters}.

\begin{figure*}
     \centering
    $$
    \begin{array}{c}
        \centering
        \includegraphics[width=1.5\columnwidth]{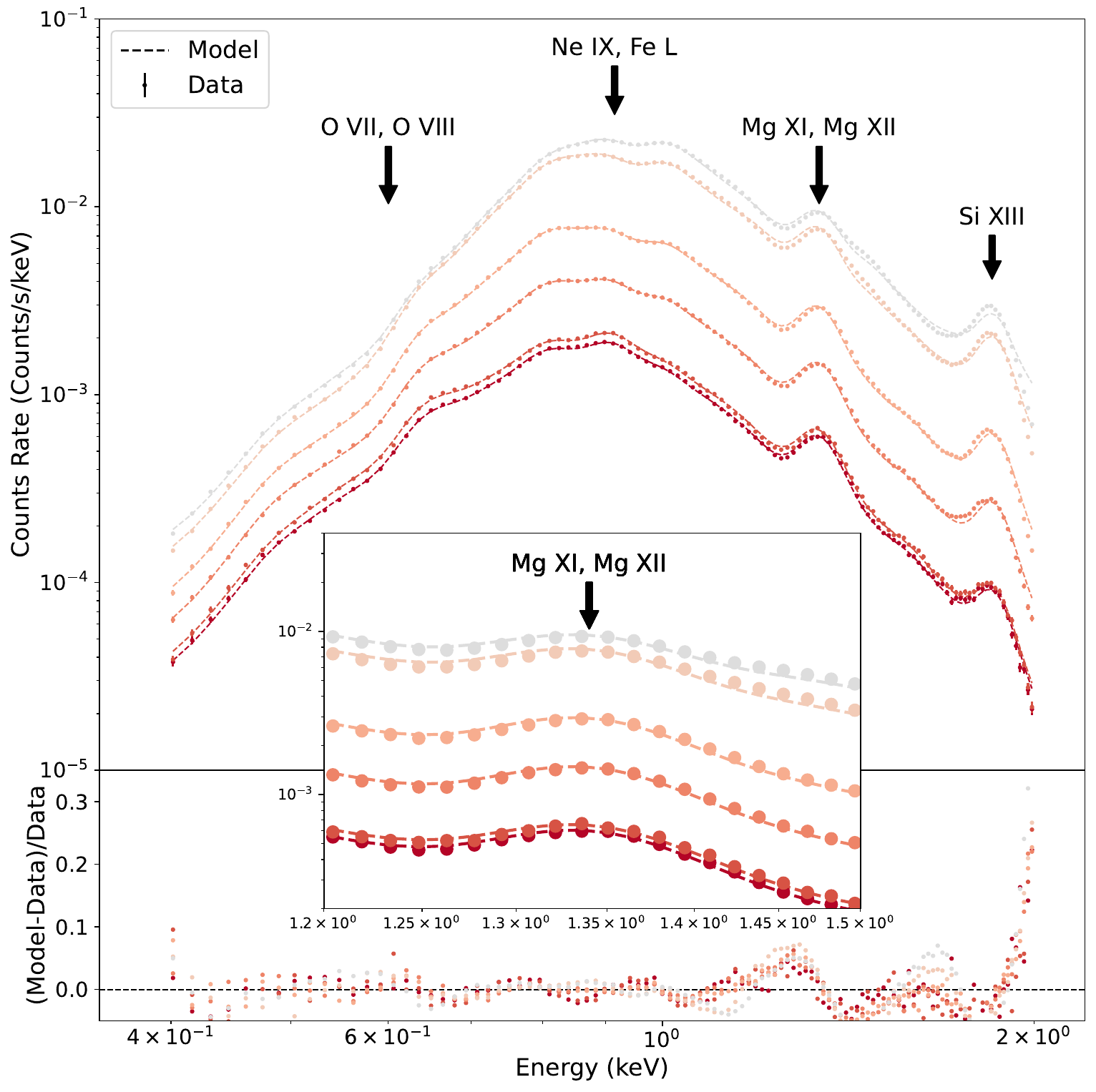} 
    \end{array}
    $$
    \caption{Mock spectra generated from \texttt{pyXSIM} and the fit from \texttt{Sherpa}. The top panel shows synthetic spectra (points with error bars) and best-fitting models (dashed lines) for the regions shown in \autoref{fig:5plots}.
    Colors indicate the region over which the spectrum was computed, starting with `D' (grey, top) and continuing from `A1' to `A5' (top to bottom, lighter to darker red, matching the colors of the rectangles in the bottom panel of \autoref{fig:5plots}). We omit the results for regions `B1' to `B5' for aesthetics, but these are qualitatively similar to the corresponding `A' regions. Labels indicate the locations of particular spectral features and the inset panel zooms in on the region from 1.2 to 1.5 keV. The bottom panel shows residuals between the data and the best-fitting model.}       
    \label{fig:create_spec_pha}
\end{figure*}

\autoref{fig:soxs_hot} shows a synthetic X-ray image for the same snapshot as \autoref{fig:5plots} as it would be observed by \textit{Chandra} using the ACIS-I configuration. The axes represent the RA and Dec for a galaxy arbitrarily located at [$30^\circ$, $45^\circ$], while the color scale indicates photon counts per pixel integrated over energies from $0.4$ to $2.0$ keV, computed using the 10 Ms results. We plot the spectra of the regions D and A1-5, again for the example case of the 110 Myr snapshot with $Z_\mathrm{bg}=0.5Z_\odot$, in \autoref{fig:create_spec_pha}.

\begin{table}
\centering
\begin{tabularx}{\columnwidth}{XX}
\hline\hline
Parameter            & Value         \\
\hline
Emission Model            & apec         \\ 
Energy Range (keV)   & 0.4 - 2.0                          \\ 
Exposure Time (Ms)   & 10        \\ 
Collecting Area (cm\(^2\)) & $\pi \times 60^2$ \\ 
Distance (kpc)       & 500                        \\ 
Sky Position (RA, Dec) & [30., 45.]          \\ 
$N_{\rm H}$ (${\rm cm}^{-2}$)               & None \\
\hline\hline
\end{tabularx}
\caption{Parameters for the \texttt{pyXSIM} package. From top to bottom, these are the emission model, energy range, exposure time, area to collect the photons, distance of the object, RA and Dec of the pointing in degrees, and absorption column. }
\label{table:pyXSIM_parameters} 
\end{table}

\begin{table}
\centering
\begin{tabularx}{\columnwidth}{XX}
\hline\hline
Parameter            & Value         \\
\hline
Energy Range (keV)   & 0.4 - 2.0                          \\ 
Exposure Time (Ms)   & 10, 1, 0.1, and 0.01                          \\ 
Sky Position (RA, Dec) & [30., 45.]          \\ 
Instrument               & chandra\_acisi\_cy22 \\
\hline\hline
\end{tabularx}
\caption{Parameters for the \texttt{SOXS} package. From top to bottom, these are the energy range, exposure time, RA and Dec of the pointing in degrees, and instrument specification. }
\label{table:SOXS_parameters} 
\end{table}

\subsection{Fitting the X-ray spectrum}
\label{sec:fitting_xray_spectrum}

The final step in our post-processing pipeline is to fit the synthetic spectra from $0.4 - 2$ keV using \texttt{SHERPA}, following the procedure commonly used for observations \citep[e.g.,][]{lopez20}. We carry out this fit assuming that the emitting medium can be described as a superposition of thermal plasmas, each characterised by a distinct temperature but with shared elemental abundances. The standard observational practice is to use one, two, or three temperature components in such fits (though of course, this is an oversimplification since the outflow is multi-phase and follows a smooth distribution in temperature -- \citealt{Vijayan&Li}). For each of these possibilities, we use \texttt{SHERPA} to find the model that best fits the spectrum in each region, as characterised by the minimum $\chi^2$ value, using \texttt{SHERPA}'s built-in Levenberg-Marquardt fitting algorithm. When carrying out these fits we treat the abundances of the elements O, Ne, Mg, Si, S, and Fe as free parameters; although we know that our model galaxies have Solar-scaled abundances, we do not add this as a constraint to the fits in order to make the fitting procedure as realistic as possible. This means that for an $N$ temperature model, we have a total of $6 + 2N$ free parameters to be fit: the abundances of six elements plus the temperature and normalisation of each component.

Due to the high-dimensional nature of the parameter space, we find that for certain values of the initial guess the fitter sometimes fails to find that true $\chi^2$ minimum; this problem is particularly likely to occur for regions far from the disc where the metallicity and SNR are lower. To mitigate this we carry out fits using a range of guesses for the initial elemental abundances from $Z/Z_\mathrm{X,\odot} = \max(Z_\mathrm{bg} - 0.5, 0)$ to $Z/Z_\mathrm{X,\odot} = Z_\mathrm{bg} + 0.5$ in steps of 0.1, where $Z_{\mathrm{X},\odot}$ is the abundance of each species X at Solar metallicity. We then accept whichever fit returns the lowest $\chi^2$ value. Once we have these best-fitting values for each of the one-, two- and three-plasma models, we accept as our final fit whichever of these produces a reduced $\chi^2$ closest to unity for each region. For all the snapshots and values of $Z_\mathrm{bg}$ we consider in this paper, we find that for the central region `D' the best fit is a three-plasma model, while all other regions are best fit by two-plasma models.

We present the best-fitting spectra for our example case as dotted lines in \autoref{fig:create_spec_pha}. The bottom panel shows the residuals (i.e., $(\mathrm{Model}-\mathrm{Data})/\mathrm{Data})$. The residuals are smaller for the lower end of the energy range ($\sim 0.5$ keV) and for regions further away from the midplane. Overall, however, the fits reproduce most aspects of the mock spectra well; the inset in \autoref{fig:create_spec_pha} demonstrates this by zooming in on the spectral region $1.2-1.5$ keV.

\section{Results and Analysis}
\label{Results and Analysis}

We are now in a position to discuss our main results. We use the procedures outlined in \autoref{sec:methods} to derive the metallicity and temperature values from the mock observations. As noted above, we carry out this analysis for $Z_\mathrm{bg} = 0$, 0.2, 0.5, 1, and 2; we also analyse a control run in which we leave the density and temperature structure unchanged, but set the metallicity to a \textit{uniform} value $Z = Z_\odot$ everywhere in the simulation domain. Except where otherwise noted we use the simulated observations for a 10 Ms exposure so that we have essentially noise-free spectra.

\subsection{Time variation}

To ensure that our findings are not dependent on the specific simulation time we choose to analyse, we use the pipeline described in \autoref{sec:methods} to process four snapshots, corresponding to the state of the simulations at 100, 105, 110, and 115 Myr. Our goal here is simply to assess the level of fluctuations in the spectra once the wind has reached a statistically steady state. We choose these specific times because the outflows have reached statistical quasi-steady-state after 100 Myr, and a 5 Myr interval between snapshots ensures that the results are not causally linked: the typical hot gas speed in the simulation is $\sim 1000$~km~s$^{-1}$, so 5 Myr interval is long enough hot gas generated at the midplane to exit the simulation domain at $\pm 4$ kpc. 

\begin{figure}
     \centering
    $$
    \begin{array}{c}
        \centering
        \includegraphics[width=\columnwidth]{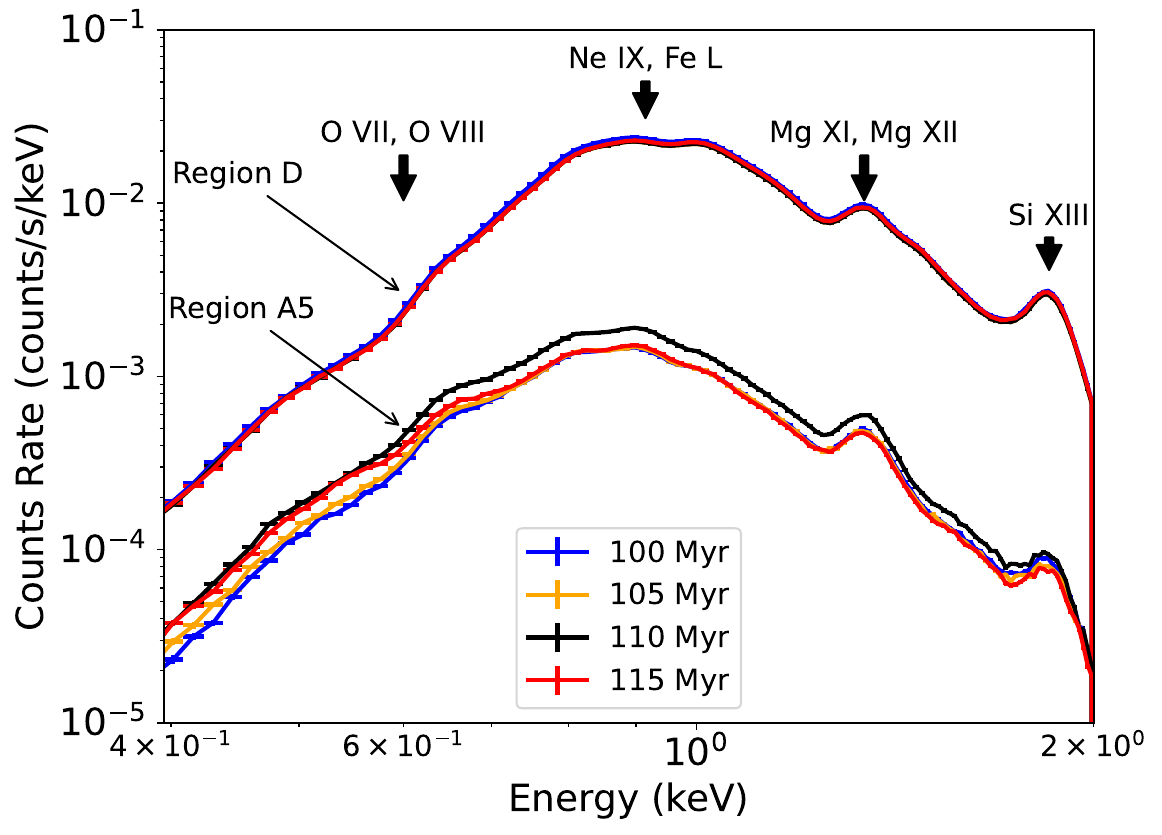} 
    \end{array}
    $$
    \caption{Synthetic spectra at region `D' and `A5' from the snapshots at 100, 105, 110, and 115 Myr, as indicated by different colors (see legend). The group of curves at the higher count rate corresponds to region `D', while the group at the lower count rate corresponds to region `A5'. Because all snapshots produce similar spectra, we focus on the at 110 Myr and generalise our results.}       
    \label{fig:spectra_all_snapshots}
\end{figure}

In \autoref{fig:spectra_all_snapshots}, we present the spectra in regions `D' and `A5' from all four snapshots. The `D' spectra are nearly identical across snapshots, while the `A5' spectra vary slightly in normalisation but show a nearly constant shape. This indicates that the outflow properties and the resultant spectra are stable over time. Quantitatively, fitting spectra at the four times shown yields best-fit estimates for elemental abundances that vary by an amount comparable to the formal uncertainties on the fits. Given this consistency, we focus our detailed analysis on the snapshot at 110~Myr for the remainder of this study, since the results are qualitatively -- and mostly quantitatively -- identical to other snapshots.

\subsection{Outflow metallicity patterns}
\label{Metallicity Gradients in Galactic Outflows}

\begin{figure}
    \centering
    $$
    \begin{array}{c}
    \centering
    \includegraphics[width=\columnwidth]{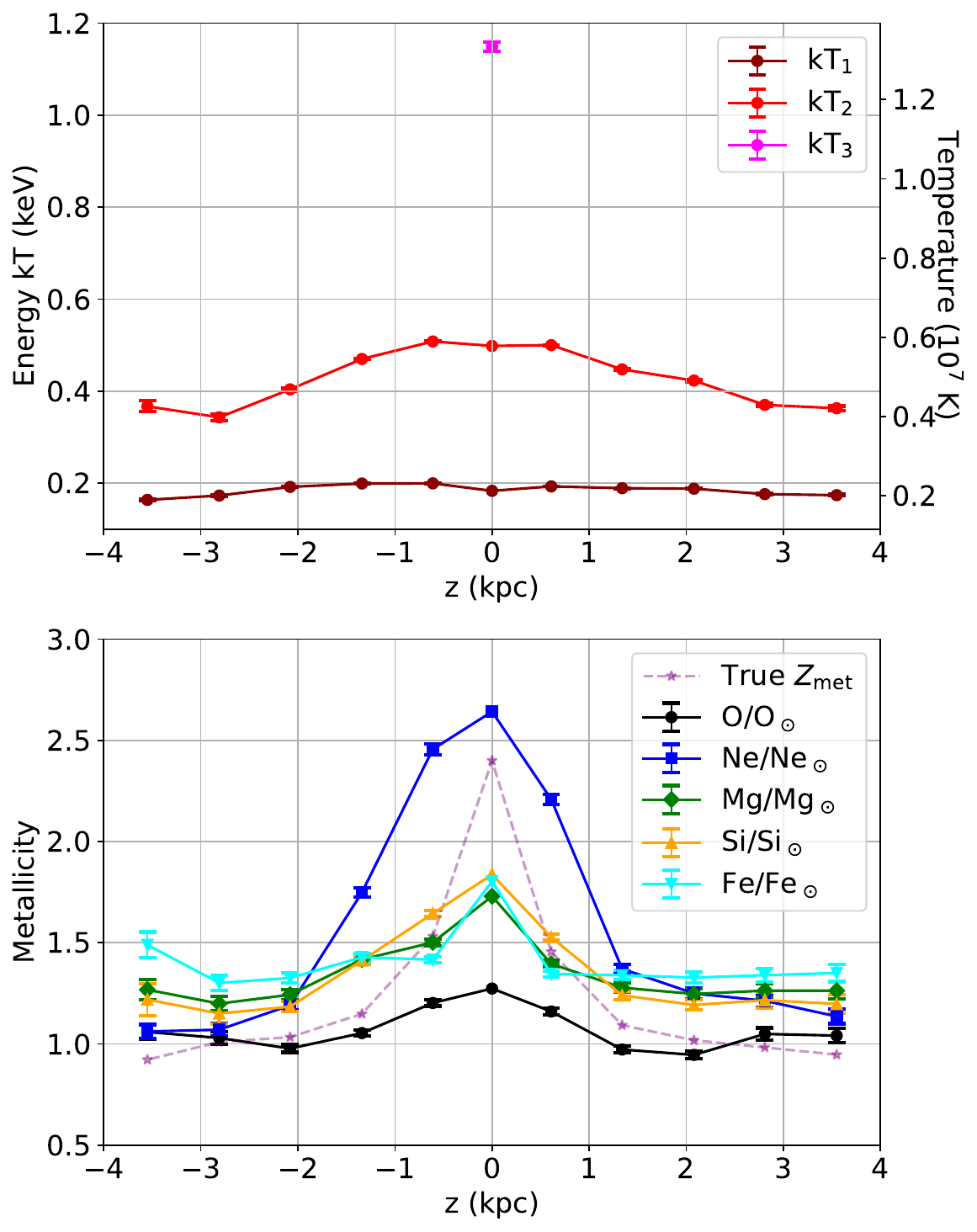} 
    \end{array}
    $$
    \caption{Top: best-fitting temperature of each plasma component as a function of distance from the midplane in the 110 Myr snapshot with $Z_\mathrm{bg} = 0.5Z_\odot$. Midplane distances are measured to the centre of each slice, and note that a third temperature component ($T_3$) is present only in the `D' region centred on $z=0$. Bottom: best-fitting abundances of the elements O, Ne, Mg, Si, and Fe normalised to Solar (solid lines, as indicated in the legend) and the true metallicity (dashed line, see \autoref{eq:Ztrue}) as a function of height.}   
    \label{fig:sherpa_hot}
\end{figure}

We first focus on the case $Z_\mathrm{bg} = 0.5$, as a baseline and to guide the discussion that follows. \autoref{fig:sherpa_hot} shows the best-fit temperature and abundances as a function of distance from the midplane for this case. We see that this fit includes one low-temperature component with $kT\approx 0.2$ keV almost independent of height, and one higher-temperature component that goes from $kT\approx 0.5$ keV at $z=0$ to $\approx 0.35$ keV at $z\approx 3.5$ kpc; in the disc region only there is also evidence for a third even hotter component at $kT \approx 1.2$ keV. The general trend that the plasma temperature falls with distance off the midplane is consistent with the phase distribution of hot gas seen in \citetalias{vijayan24} and shown in \autoref{fig:5plots}. 

The abundances of O, Ne, Mg, Si, and Fe plotted in the bottom panel of \autoref{fig:sherpa_hot} show similar patterns, which is expected given that we have assumed Solar-scaled abundances and thus the true abundance ratios are constant. This pattern is that the outflowing gas possesses significant metallicity gradients, with a peak in the central region D and a subsequent decline for A/B-1-5. Such a trend in metallicity is strikingly similar to that observed by \citet{lopez20} for S and Si in the wind of M82. We discuss the physical origin of this pattern and its implications in \autoref{Discussion}.

For comparison, we also show the ``true'' hot gas metallicity, reminding readers that, because our simulation assumes Solar-scaled abundances, the true (Solar-normalised) metallicity is the same for all elements. There is some ambiguity about how to define this, since different choices of how to average over the metallicity distribution present in the real simulation data yield different outcomes. In order to produce something as close as possible to the quantity returned by the observations, we define our true metallicity gradient to be the X-ray luminosity-weighted mean metallicity. That is, we define the true metallicity as
\begin{equation}
    Z_\mathrm{met} = \frac{\int \epsilon_X(Z,T) Z\,dV}{\int \epsilon_X(Z,T) \, dV},
    \label{eq:Ztrue}
\end{equation}
where $T$ and $Z = Z_\mathrm{bg} + Z_\mathrm{SN}$ are the temperature and total metallicity in each cell, $Z_\mathrm{SN}$ is the metallicity contributed by SNe as measured from the concentration of passive scalar in the simulation, $\epsilon_X(Z,T)$ is the luminosity per unit volume from 0.4 - 2 keV (as shown in the fourth row of \autoref{fig:5plots}), and the integral goes over all the cells that belong to a given slice. We see that none of the observationally-inferred gradients quite reproduce the true metallicity structure, shown in the dashed line in \autoref{fig:sherpa_hot}, and that the fits show an asymmetry between the regions above and below the midplane that is not present in the underlying data. This is perhaps not surprising: the fits rely on the hugely simplifying assumption that there are only a small number of single-temperature plasma components, each with the same abundances. Neither of these assumptions are true, and given the very high SNR synthetic spectra we have used to carry out these fits, we expect these systematic errors to dominate the error budget. Nonetheless, the qualitative trend we find in the simulation data -- a maximum in the metallicity in the disc region and a decrease away from it -- is reproduced for each fitted element.

\begin{figure}
\centering
\includegraphics[width=\columnwidth]{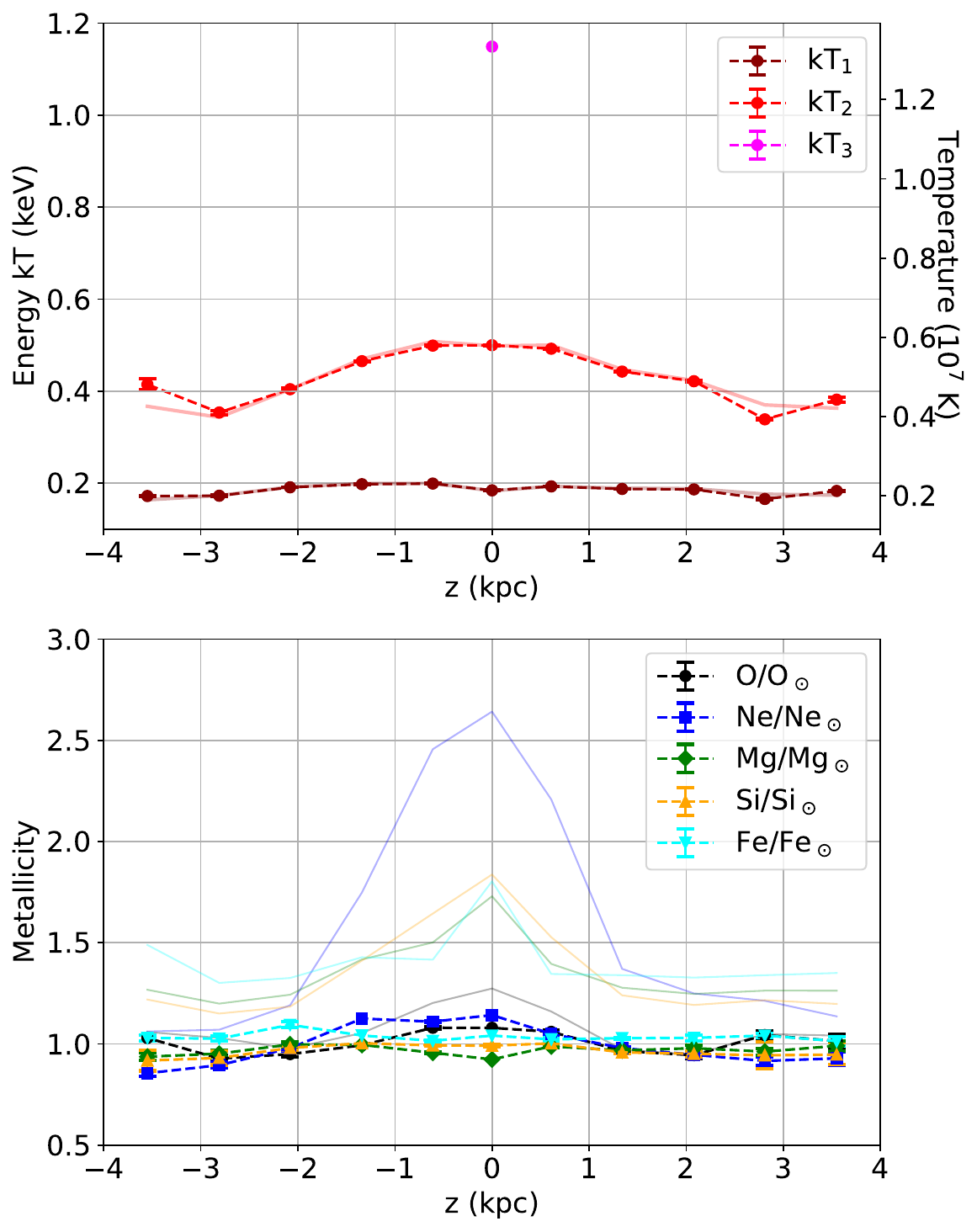}
\caption{Same as \autoref{fig:sherpa_hot}, but now comparing the standard case $Z_\mathrm{bg} = 0.5$ to a control run where we set $Z=Z_\odot$ in all cells. In the plots above, dash lines with points and error bars are the results for the control run, while faded lines are the $Z_\mathrm{bg} = 0.5$ case for comparison; these lines are identical to those shown in \autoref{fig:sherpa_hot}.}
\label{fig:control_run}
\end{figure}

As a further confirmation that the fits are responding to the real metallicity gradient present in the data, and are not simply an artefact introduced by the temperature gradient, we can compare to the control run which has $Z=Z_\odot$ everywhere. We show the results for this case superimposed on the results for $Z_\mathrm{bg}=0.5$ in \autoref{fig:control_run}. We see that in the control run the best-fitting temperatures are nearly identical, as expected, but we now recover no gradient in the abundances. Thus our fits correctly diagnose this case as having a wind of uniform metallicity. This gives us confidence that, while the inferred metal abundances do not precisely recover the true distribution, they do capture the major qualitative feature of the real distribution and are unlikely to be false positives.

\subsection{Results for varying background metallicity}
\label{Impact of Different Background Metallicity}

\begin{figure}
\centering
\includegraphics[width=\columnwidth]{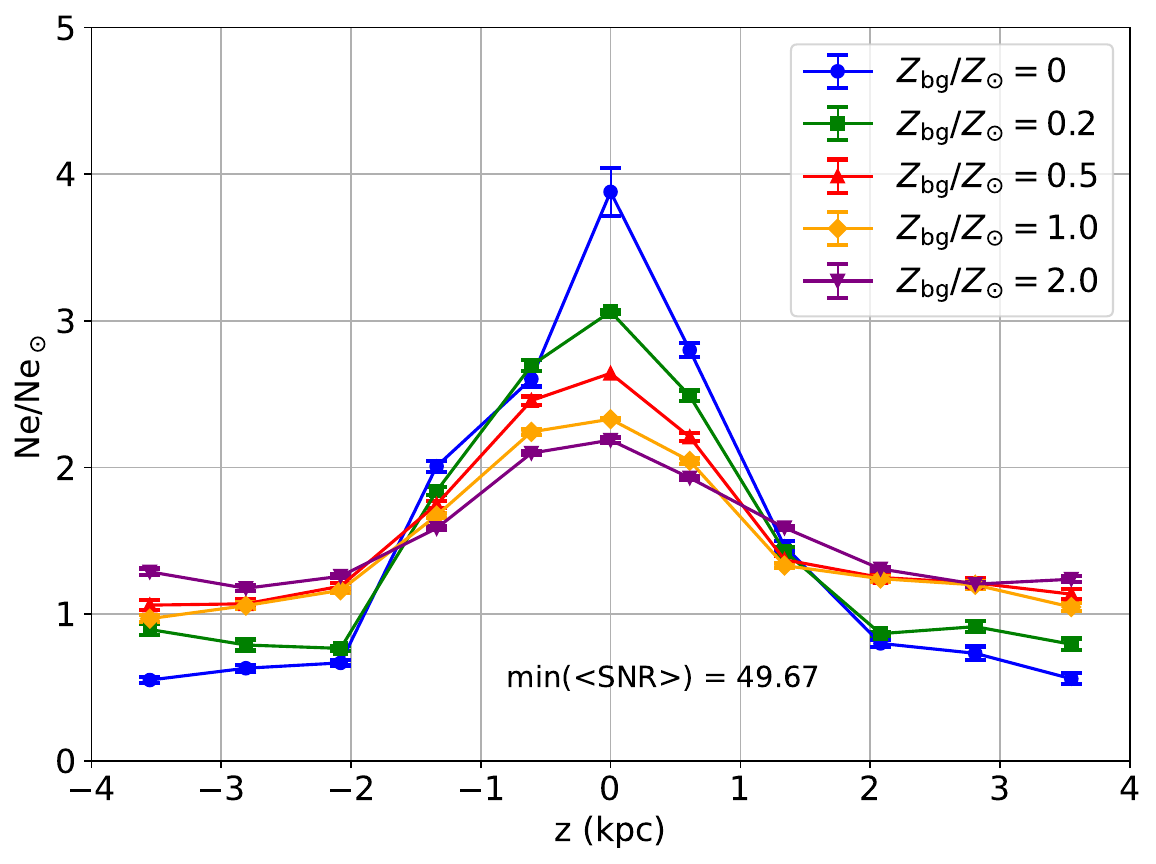} \\
\includegraphics[width=\columnwidth]{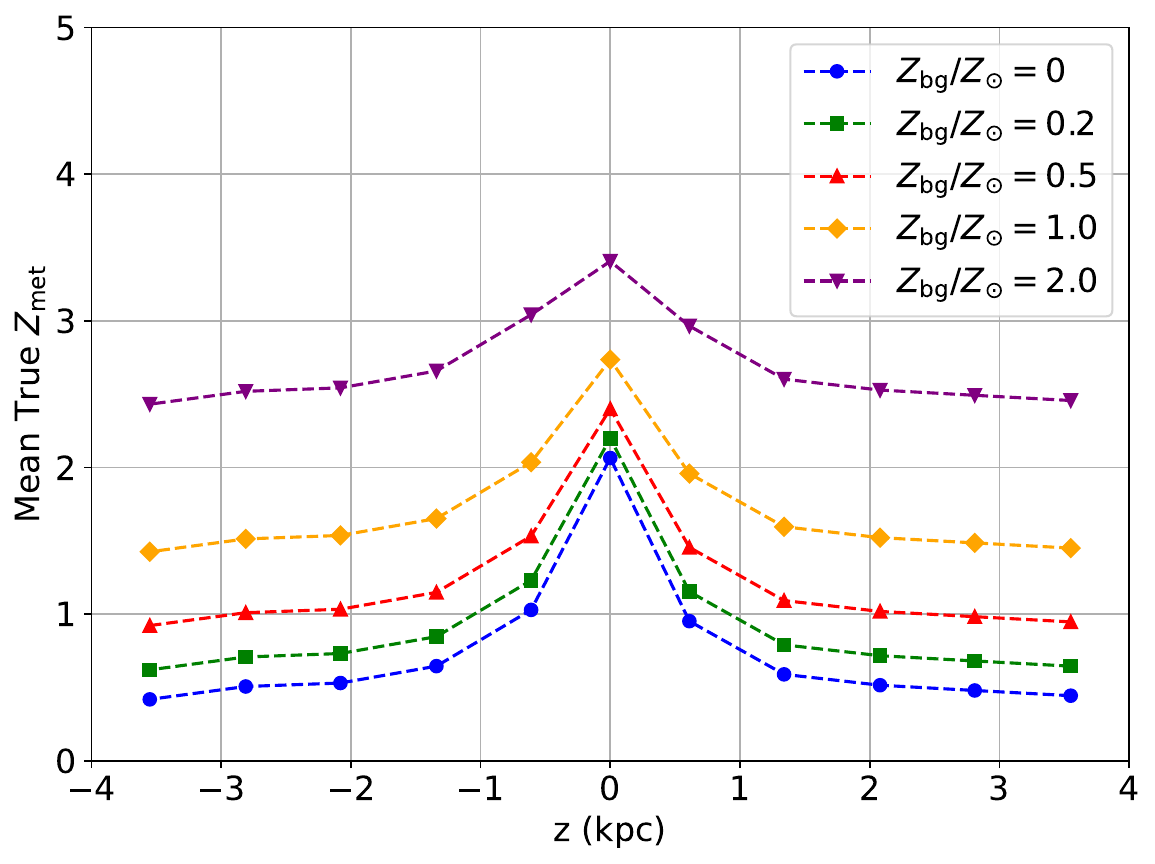}
\caption{Top: best-fit Ne abundance as a function of height off the midplane for varying levels of background metallicity $Z_\mathrm{bg}$, as indicated in the legend. Bottom: true metallicity (\autoref{eq:Ztrue}) as a function of height estimated using different $Z_{\rm bg}$.}
\label{fig:metal_gradient_Ne}
\end{figure}

We now expand our results to other $Z_{\rm bg}$ values to understand how the observationally-inferred metallicity gradient changes as we vary the strength of the background metallicity upon which the SN-injected metals are imposed. We show the result of this experiment for Ne in the top panel of \autoref{fig:metal_gradient_Ne}; we choose Ne because it is the element that shows the steepest gradient in our fits with $Z_\mathrm{bg} = 0.5$, but the qualitative result is the same for all elements. By contrast, we show the true X-ray luminosity-weighted mean metallicities for these runs in the bottom panel.

We retrieve vertical gradients in metallicity for all values of $Z_{\rm bg}$. The pristine case of $Z_{\rm bg}=0$ shows the strongest gradient in inferred Ne abundance, with a factor of $\sim 8$ difference between the abundances recovered at $z=0$ and $z\approx |4|$ kpc. Gradients become successively flatter with increasing $Z_{\rm bg}$, with the case $Z_\mathrm{bg} = 2$ showing less than a factor of two difference between regions D and A5 or B5. Compared to the true gradients, we see that the numerical values do not match precisely, but the sense of variation does, i.e., as we reduce $Z_\mathrm{bg}$ both the true gradient and the gradient in inferred Ne abundance become steeper. However, we also note some surprising behaviour: the retrieved central abundance \textit{decreases} with increasing $Z_\mathrm{bg}$, opposite to the trend in the actual metallicity (cf. the top and bottom panels of \autoref{fig:metal_gradient_Ne}).

We can understand this behaviour as arising from a fundamental limitation of the commonly-used observational fitting procedure, which is that it assumes that the different temperature components share a common set of abundances. This assumption, while it is likely a practical necessity to avoid having the number of free parameters in the fit balloon out of control, is fundamentally incorrect, particularly near the disc, because the emitting medium here consists largely of bubbles of near-pure SN ejecta that are both very hot ($\sim 10^7$ K) and very metal-rich, and a slightly cooler (but still $\sim 10^6$ K) component of shocked ISM gas that is more metal-poor. Now let us consider the implications of this configuration for how the X-ray spectrum changes as we vary $Z_\mathrm{bg}$. The bubbles of SN-ejecta produce emission that is not tremendously sensitive to $Z_\mathrm{bg}$, because their metallicities are dominated by direct SN ejecta; they will be metal-rich even for $Z_\mathrm{bg} = 0$. The shocked ISM component, on the other hand, has a luminosity that is highly sensitive to $Z_\mathrm{bg}$; low $Z_\mathrm{bg}$ makes this component dimmer compared to the emission from the fresh SN ejecta, while higher $Z_\mathrm{bg}$ makes it brighter. We can therefore understand the effect of $Z_\mathrm{bg}$ as changing the weight of the two components, each of very different intrinsic metallicity, that contributes to the overall spectrum.

When we fit this composite spectrum with a model that assumes that there is a single metallicity, the fit naturally tries to find a compromise between the two components. However, it winds up favouring whichever one is brighter. Thus when $Z_\mathrm{bg}$ is small and emission is dominated by the fresh SN ejecta, the fit returns a very high metallicity characteristic of those ejecta. As we increase $Z_\mathrm{bg}$ and thus the relative contribution from the shocked ISM component, the fit moves toward the lower metallicity of this component, explaining why the inferred abundances are lower at higher $Z_\mathrm{bg}$.

To confirm this hypothesis, we have experimented with generating spectra with emission from the most metal-rich gas, $Z > 4 Z_\odot$, artificially set to zero. We find that, when we do this, the anomalous behaviour that inferred abundances decrease with increasing $Z_\mathrm{bg}$ vanishes. Instead, consistent with our proposed hypothesis, we recover abundances that increase with $Z_\mathrm{bg}$ as expected.

In regions further away from the midplane ($|z|>2$ kpc) this effect vanishes and increasing background metallicity increases spectrum-predicted metallicity. This is because at these greater distances from the injection point, the super-hot phase has now mixed and cooled down and the dominant X-ray emitting gas is at $\sim 10^6$ K. We no longer have emission coming from two phases at so starkly different metallicities and temperatures, but are closer to the state envisioned by the fit where there is a range of temperatures but only a single metallicity. The better match between the true and assumed gas metallicity and temperature distributions in turn means that the fit works better, and we no longer have the erroneous result that higher $Z_\mathrm{bg}$ pushes the fit in the wrong direction. 

Finally, we note that neither the ``true'' X-ray-weighted metallicity nor the elemental abundances we retrieve from the X-ray spectrum fully agree with the mass-weighted metallicities computed directly from the simulation data, even if we limit ourselves to the hot phase. In particular, Figure 9 of \citetalias{vijayan24} shows that for the hot phase, the mass-weighted metallicity is near-identical to the background metallicity for $Z_{\rm bg}\gtrsim 0.2$. Thus weighting by X-ray luminosity or the emission spectrum tends to accentuate the gradients compared to the mass-weighted average. For the true metallicity values, the averaging is biased towards regions of higher density, particularly because the luminosity is proportional to the square of the density. Therefore, even a small increase in emissivity due to increased background metallicity disproportionately biases the true metallicity average towards gas with a higher background, making the inferred metallicity appear higher than a mass-weighted value would suggest. We explore this issue further in \autoref{sec:What Biases in X-ray Derived Abundances?}.


\subsection{Impact of finite signal-to-noise ratio}
\label{sec:impact_of_snr}

Thus far we have restricted our analysis to spectra generated for an exposure time of 10 Ms, so we are examining essentially noise-free data. It is therefore important to consider the more realistic case of observations with a finite signal-to-noise ratio (SNR), in order to determine what quality data are required to retrieve the essential features of the outflow metallicity pattern. 

To quantify spectral quality, we define the mean photon count-weighted $\langle \mathrm{SNR}\rangle$ to average over the SNRs in each energy bin:
\begin{equation}
\langle \mathrm{SNR}\rangle = \frac{\sum\limits_i c_i \cdot \text{SNR}_i}{\sum\limits_i c_i}\,, 
\end{equation}
where the sum runs over all energy bins from $0.4-2$ keV, $c_i$ is the photon count in bin $i$ (which is proportional to the exposure time), and $\text{SNR}_i$ is the SNR of bin $i$. Assuming that we are dominated by Poisson noise, $\text{SNR}_i = c_i^{1/2}$, and this gives
\begin{equation}
\langle \mathrm{SNR}\rangle = \frac{\sum\limits_i c_i^{\frac{3}{2}}}{\sum\limits_i c_i}.
\label{eq:SNR}
\end{equation}
\autoref{eq:SNR} provides us with a means of calculating the mean SNR of the spectrum in each region. To characterise the overall SNR of an observation, we then take the minimum over all regions, i.e., we define
\begin{equation}
    \mathrm{SNR}_\mathrm{obs} = \min_j \left(\langle\mathrm{SNR}\rangle_j\right),
    \label{eq:snr_obs}
\end{equation}
where $j$ runs over our observing regions, i.e., we let $j = \mathrm{A5}, \mathrm{A4}, \ldots, \mathrm{D}, \mathrm{B1},\ldots \mathrm{B5}$.

\begin{figure*}
     \centering
    $$
    \begin{array}{cc}
    \centering
    \includegraphics[width=\columnwidth]{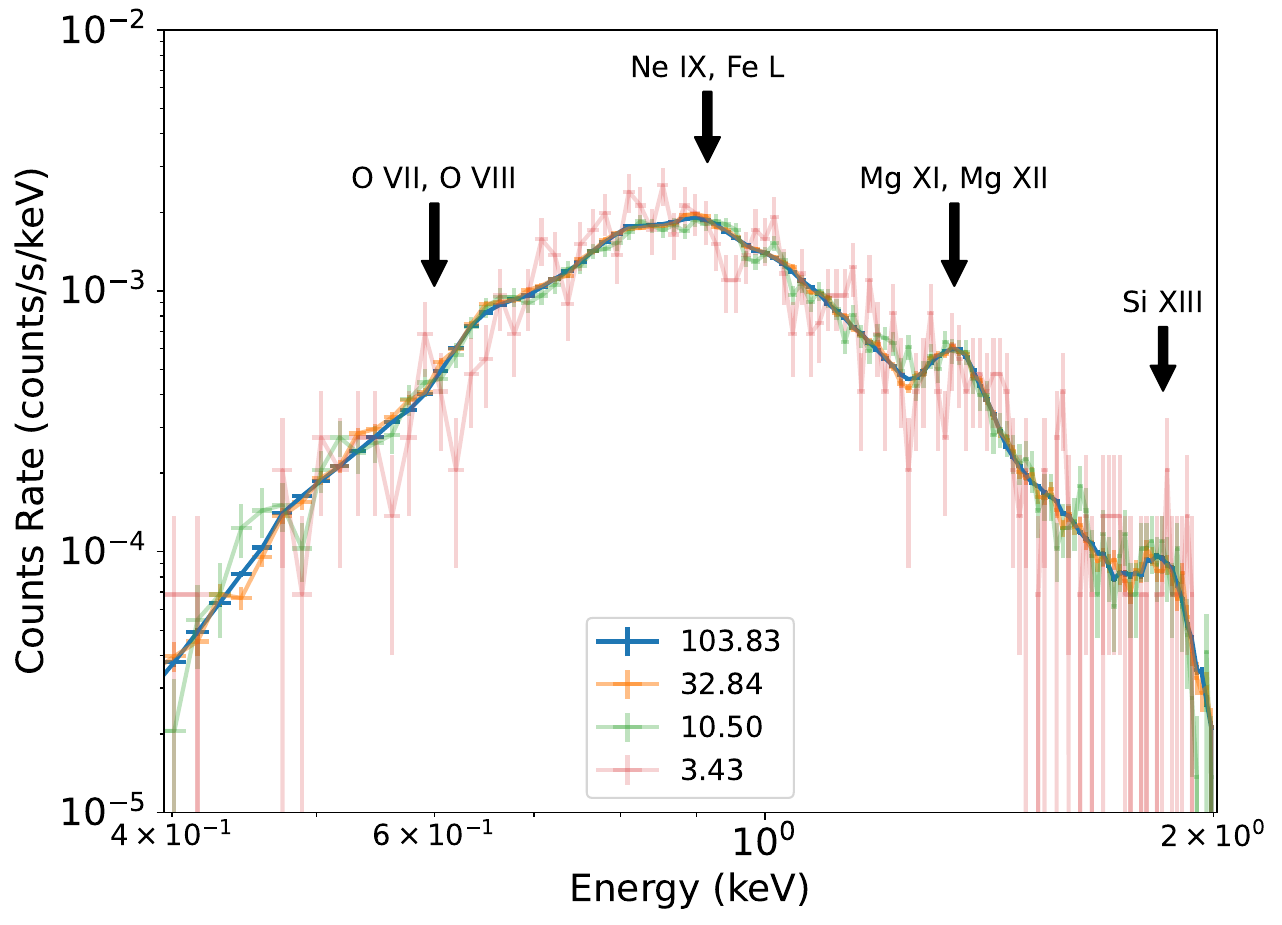} & 
    \includegraphics[width=\columnwidth]{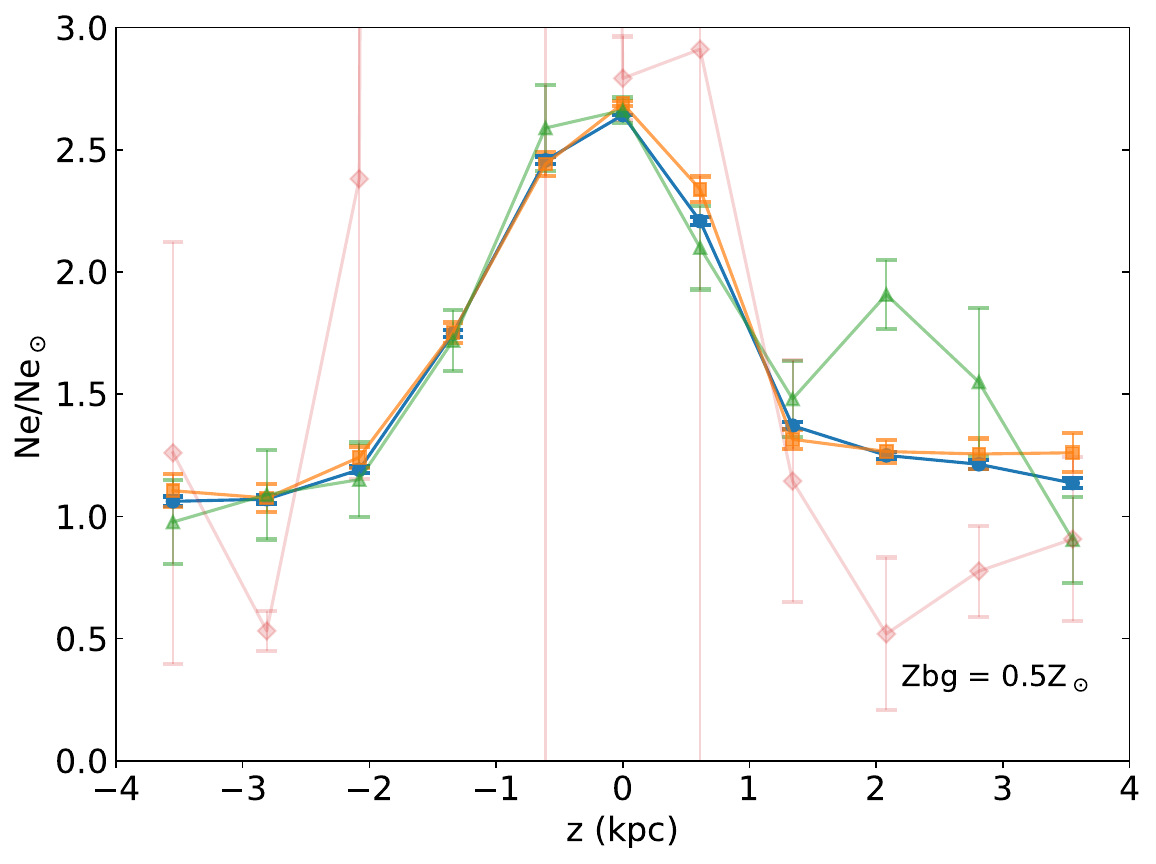}
    \end{array}
    $$
    \caption{Left: spectra for region `A5'  for the case of $Z_{\text{bg}} = 0.5Z_{\odot}$ at 110 Myr measured at four different exposure times -- 10, 1, 0.1, and 0.01 Ms. The legend indicates the value of $\mathrm{SNR}_\mathrm{obs}$, the overall SNR we estimate from the exposure time (\autoref{eq:snr_obs}). Right: Ne abundances as a function of height recovered from spectral fitting, for the same four SNRs shown in the left panel.}
    \label{fig:impact of SNR}
\end{figure*}

To generate results for a range of values of $\mathrm{SNR}_\mathrm{obs}$, we vary the exposure time from the initial 10 Ms, decreasing it to 1, 0.1, and 0.01 Ms. For each of these cases we produce mock spectra with \texttt{SOXS} and fit using \texttt{Sherpa}, and we plot the resulting spectra for the A5 region and the spectrum-derived Ne abundance for all regions in the right and left panels of \autoref{fig:impact of SNR}, respectively; we again focus on Ne as an example, but the qualitative results are the same for other elements. In both the panels, the curves are labeled by $\mathrm{SNR}_\mathrm{obs}$. The blue curve ($\mathrm{SNR}_\mathrm{obs}=103.83$) is the case for a 10 Ms exposure, and is identical to the red curve in \autoref{fig:metal_gradient_Ne}; exposure times of 1, 0.1, and 0.01 Ms yield $\mathrm{SNR}_\mathrm{obs}=32.84$, 10.50, and 3.43, respectively, as indicated in the legend. As one might expect, for poor SNRs, the deviation from the ``perfect'' spectrum is smallest for mid-energy bins ($E\sim 1$ keV) where \textit{Chandra} is most sensitive, and increases rapidly for lower and higher energy bins. This may be the result of the relatively poor sensitivity of \textit{Chandra} at these energies. 

From the right panel, we see that the retrieved abundances begin to deviate from the ``perfect data'' case as the SNR degrades. At a SNR of 33, the results match the perfect case to within fit uncertainties, and even for a SNR of 10.5 the main feature of the metal distribution -- in particular that there is a maximum near $z=0$ and a decrease away from it -- remains visible, albeit with an additional spurious peak appearing on the $+z$ side. Interestingly, this spurious peak looks quite similar to a feature that \citet{lopez20} detected for Ne in their analysis of the M82 wind -- cf.~their Figure 5. For a SNR of 3.4, even broad features of the abundance distribution are lost, and in fact the fit fails to converge at all for the bin at $z\approx -1$ kpc. This suggests that X-ray observations seeking to retrieve metal abundances in galactic outflows should aim for a minimum mean SNR of at least 10, and preferably closer to 30. With this level of SNR, the results will be dominated by the systematic errors coming from the simplified model of the emitting plasma rather than the random errors arising from photon counting statistics.

\section{Discussion}
\label{Discussion}

Here we discuss two implications of our work: first that negative metallicity gradients in galactic outflows provide unambiguous observational evidence for metal loading (\autoref{sec:Implementation of Metallicity Gradients}), and second that measurements of such gradients can be, at least approximately, turned into quantitative measurements of phase mixing in galactic outflows (\autoref{sec:analytical_model}).

\subsection{Negative metallicity gradients as evidence for metal loading}
\label{sec:Implementation of Metallicity Gradients}

Our findings offer a clear explanation for the observation that galactic winds show $\alpha$ abundances that are enhanced relative to those of the ISM, but that this enhancement declines away from the galactic plane. This pattern occurs due to the distribution of metals among different gas phases and their respective contributions to X-ray emission. In the disc of the galaxy, recent supernova activity produces very hot gas that is highly enriched with $\alpha$ elements, leading to X-ray emission characteristic of elevated metallicities. At the same location there is a cold phase ($T < 10^4$~K) of gas with much lower metallicity (but which nonetheless carries most of the total metal mass), this gas does not contribute to X-ray emission because of its low temperature, so the X-ray spectrum is dominated by the hot, very metal-rich gas from supernova ejecta. As the hot gas flows outward, however, it mixes with clouds of colder, metal-poorer gas from the surrounding ISM that have been entrained into the outflow. Over time and as the outflow moves away from the galactic plane, some of this cooler gas evaporates into the hot phase,  diluting it and decreasing the mean metallicity of the X-ray emitting gas by an amount that increases with distance from the plane. This leads to a metallicity gradient where the X-ray emitting gas near the disc has higher metallicity than that further out in the wind. Although in our simulations we have assumed Solar-scaled abundances for simplicity, in reality, this trend should occur primarily for the $\alpha$ elements that are preferentially produced by core collapse SNe. This phenomenon explains the trends observed by \citet{lopez20}.

The converse point is perhaps more interesting: as shown in \autoref{fig:control_run}, when we consider the control case of a wind that is \textit{not} heavily metal loaded and has roughly the same metallicity as the ISM, the observed gradient vanishes. This means that the abundance pattern we observe in the wind of M82 requires that unmixed SN ejecta make a significant contribution to the total wind budget. In the language commonly-used in galactic chemical evolution modeling, the observations of M82 require that the metal loading factor $\zeta$, which measures the metallicity of the wind relative to that of the ISM \citep{Peeples11a}, be larger than unity, while the yield reduction factor $\phi$, which parameterises the fraction of newly-produced metals from SNe that are retained in the disc rather than lost promptly to an outflow \citep{Sharda21a}, must be significantly below unity. 

\subsection{Analytical Model of Phase Mixing}
\label{sec:analytical_model}
\begin{figure}
     \centering
    $$
    \begin{array}{c}
    \centering
    \includegraphics[width=\columnwidth]{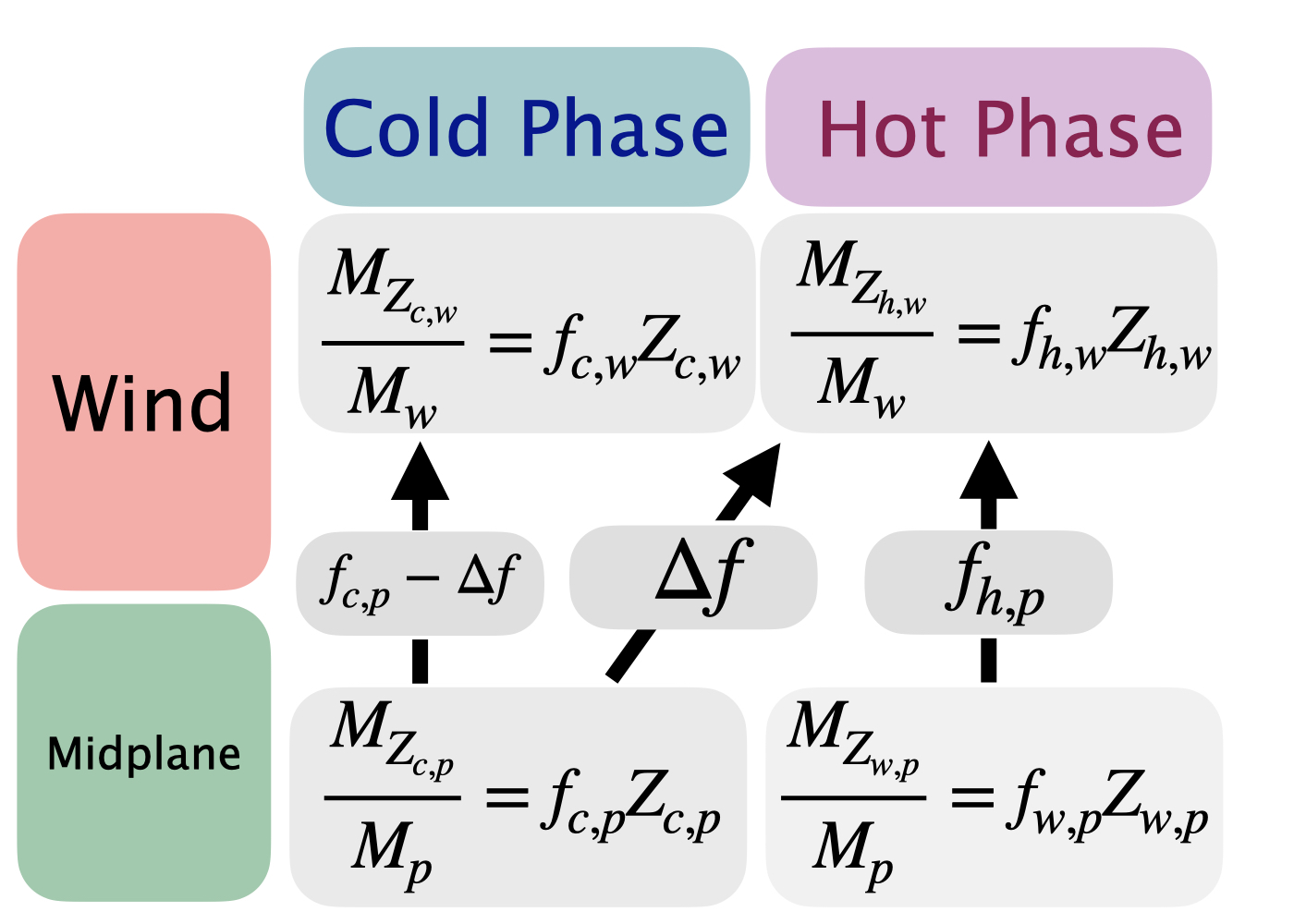}
    \end{array}
    $$
    \caption{Visualization of the `4-grid model'. This schematic illustrates the division of gas into midplane and wind zones, each containing cold and hot phases. The model represents the transfer of mass fraction between these components, with $\Delta f$ indicating the net increase in hot gas fraction from the midplane to the wind. Arrows represent potential gas flows and mixing processes between the components.}
    \label{fig:4-grid model}
\end{figure}

The explanation of abundance gradients that we have just presented raises an obvious question: can we use measured metallicity gradients to deduce the nature of phase mixing in galactic winds? To see that this is possible, it is helpful to introduce a toy `4-grid model' that quantifies the extent of mixing between the hot ($T>10^{6}$ K) and cold ($T<10^{4}$ K) phases of the outflows from observations of gradients; we illustrate this model schematically in \autoref{fig:4-grid model}.

We divide the gas into two zones -- the midplane and the wind -- with a boundary at $\sim \pm 1$ kpc, roughly the height at which the outflows are launched. Each zone contains both hot and cold components, giving us four ``boxes'' of gas: cold gas in the plane (c,p), hot gas in the plane (h,p), cold gas in the wind (c,w), and hot gas in the wind (h,w). The respective metal masses in these boxes are denoted by $M_{Z_\text{c/h,p/w}}$, and we can express these metal masses as the product of the total mass in each zone $M_\text{p}$ and $M_\text{w}$ times the fraction $f$ of that mass in the hot or cold phases times the mean metallicity $Z$ of each phase, i.e., $M_{Z,\text{c,p}} = Z_{\text{c,p}} f_{c,p} M_\text{p}$ and similarly for the other three gas types shown in \autoref{fig:4-grid model}. Here $f_\text{c,p}$ for example is the fraction of in-plane gas that is in the cold phase, and clearly, we have $f_{\text{c,p}} + f_{\text{h,p}} = 1$ and $f_{\text{c,w}} + f_{\text{h,w}} = 1$.

Since we have seen that evaporation of cold gas into the hot phase is the key physical process driving metallicity gradients, we let $\Delta f$ represent the net transfer of gas from the cold midplane to the hot wind, driven by the heating and evaporation of cold midplane gas into the hot wind phase. Hence, the value of $\Delta f$ is
\begin{equation}
\Delta f = f_{\text{c,p}}-f_{\text{c,w}} = f_{\text{h,w}}-f_{\text{h,p}}. 
\label{eq:Delta_f}
\end{equation}
The metal mass in the hot wind can therefore be written as a sum of two components: the metal in hot gas outflowing from the midplane and the metal added by evaporation of cold midplane gas into the hot wind phase,
\begin{equation}\label{eqn:metal_mass_hot}
\frac{M_{Z_\text{h,w}}}{M_\text{w}} = f_\text{h,p} Z_\text{h,p} + \Delta f Z_{\text{c,p}}\,.
\end{equation}
But since $M_{Z_\text{h,w}} = f_\text{h,w} Z_\text{h,w} M_w$, this means that
\begin{equation}
f_{\text{h,w}} Z_{\text{h,w}} = f_{\text{h,p}} Z_{\text{h,p}} + \Delta f Z_{\text{c,p}}\,.
\end{equation}
If we now substitute for $f_\text{h,w}$ in terms of $\Delta f$ using \autoref{eq:Delta_f}, we can rewrite this relation as
\begin{equation}
\frac{\Delta f}{f_{\text{h,w}}} = \frac{1-Z_{\text{h,w}}/Z_{\text{h,p}}}{1 - Z_{\text{c,p}}/Z_{\text{h,p}}}. 
\label{eq:true}
\end{equation}

The relation above is useful because it connects directly observable quantities, the metallicity ratio between the hot and cold gas in the central disc ($Z_{\text{h,p}}$ and $Z_\text{c,p}$) and the that in the outflowing wind ($Z_{\text{h,w}}$ and $Z_\text{c,w}$), to something that is not directly observable but is of significant interest: the fractional increase in the fraction of outlowing material that is hot due to the evaporation of cool clouds into the wind ($\Delta f/f_{\text{h,p}}$). Thus we can use the evolution of the metallicities with height as a direct probe of phase mixing.

In the limiting case where the background ISM has low metallicity, we can simplify this expression even further by assuming that $Z_{\text{c,p}} \ll Z_{\text{h,p}}$, i.e., that the hot gas freshly injected by supernovae is much, much more metal-rich than the surrounding cool ISM. In that case, our expression simplifies even further to
\begin{equation}
\frac{\Delta f}{f_{\text{h,w}}} \approx 1 - \frac{Z_{\text{h,w}}}{Z_{\text{h,p}}},
\end{equation}
and recalling that $\Delta f = f_{\text{c,p}}-f_{\text{c,w}} = f_{\text{h,w}}-f_{\text{h,p}}$ we then have
\begin{equation}
\frac{f_{\text{h,p}}}{f_{\text{h,w}}} \approx \frac{Z_{\text{h,w}}}{Z_{\text{h,p}}}.
\label{eq:approx final}
\end{equation}
In this case, the metallicities of the cold gas vanish entirely, and we can directly read off the increase in the hot gas mass fraction due to evaporation from the ratio of the hot phase metallicities in the plane versus the wind. Thus for example, if observations of a galaxy reveal that the metallicity of the hot wind is half that of the hot midplane hot gas (i.e., $Z_{\text{h,w}}/Z_{\text{h,p}} = 0.5$, roughly what \citet{lopez20} find for Si in the wind of M82), the simplified relation implies that the hot gas fraction in the midplane is approximately half that in the wind (i.e., $f_{\text{h,p}}/f_{\text{h,w}} \approx 0.5$), so the wind a few kpc from the galaxy consists roughly equally of gas that was heated at the midplane and escaped while hot, and cool gas that was entrained at the midplane but subsequently evaporated into the hot phase. However, we caution that this result will always be at least somewhat approximate, because we have seen that the elemental abundances inferred from X-ray measurements do not perfectly reflect the underlying abundances. For example, if we examine \autoref{fig:sherpa_hot}, we can see that carrying out this exercise for O versus Ne would lead to quite different numerical values for $f_{\text{h,p}}/f_{\text{h,w}}$. We explore the reasons for these errors further in \autoref{sec:What Biases in X-ray Derived Abundances?}.

\subsection{Why Do Biases in Derived Abundances Vary with Element?}
\label{sec:What Biases in X-ray Derived Abundances?}

Following our toy model of phase mixing, we now examine why X-ray derived abundances can differ systematically from the underlying metallicities, and do so differently for different elements. The key drivers of these differences are that (1) there is no single characteristic temperature for the ``hot'' ($T>10^{6}$ K) phase, but rather a broad distribution of temperature, (2) that temperature and metallicity are highly correlated, and (3) due to the different lines that contribute to the X-ray spectrum from $0.4-2$ keV, different elements are sensitive to gas at different temperatures, rather than simply reflecting some sort of mean X-ray weighted metallicity such as $Z_\mathrm{met}$ (\autoref{eq:Ztrue}).

\begin{figure}
    \centering
    \includegraphics[width=\columnwidth]{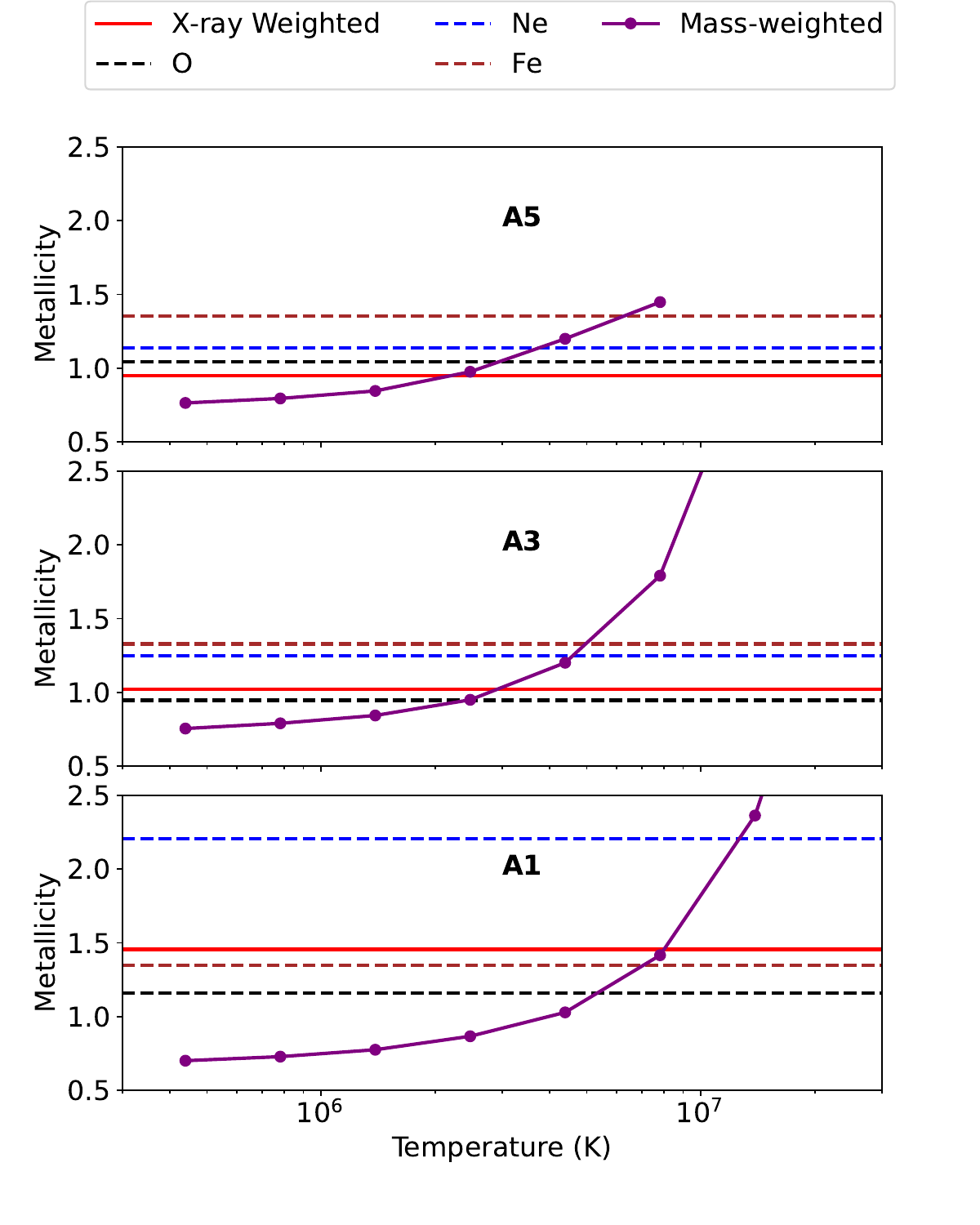}
    \caption{Mass-weighted metallicity in discrete temperature bins (purple curves), plotted together with horizontal lines showing the derived metallicities for oxygen (black dashed), neon (blue dashed), and iron (brown dashed). The red solid line shows the ``true'' X-ray luminosity-weighted metallicity $Z_\mathrm{met}$. We display three representative wind regions (`A1', `A3', `A5') in stacked panels. O~\textsc{vii} and O~\textsc{viii} lines are most sensitive to lower temperatures ($T \approx (2-3)\times10^{6}$ K), while Ne and Fe trace hotter gas, which combined with the metallicity-temperature correlation lead metallicities derived based on the O abundance to be systematically smaller than those derived from Ne or Fe. }
    \label{fig:mass_weighted_metallicity_vs_temperature}
\end{figure}

We illustrate the first two points using the case $Z_\mathrm{bg} = 0.5Z_\odot$ as an example; for each region of this case we divide the gas hotter than $10^6$ K (the temperature range that drives the bulk of the X-ray emission; cf.~\autoref{fig:frac_epsilon_vs_frac_V_&_T}) into temperature bins 0.25 dex wide, and within each bin we compute the mass-weighted mean metallicity. We show the result for regions A1, A3, and A5 in \autoref{fig:mass_weighted_metallicity_vs_temperature}; other regions are qualitatively similar. A key point to take from these plots is that, except in region A5, ``hot'' gas spans more than an order of magnitude in temperature, and metallicity is highly-correlated with temperature, particularly close to the galactic plane. The origin of this correlation is simply that one of the most important cooling channels for the very hot gas produced by SNe, particularly at small $z$ when there has been little time for adiabatic expansion or radiative cooling, is dilution by mixing with cooler gas, and dilution simultaneously cools and lowers metallicity.

To illustrate how this interacts with the third point, regarding the different temperature sensitivity of different elements, we also in \autoref{fig:mass_weighted_metallicity_vs_temperature} show the metallicities in each region inferred from the X-ray spectral fits for three sample elements, O, Ne, and Fe. We choose these three because, as shown in \autoref{fig:sherpa_hot}, O and Ne show systematically the lowest and highest inferred abundances near the midplane, while Fe is intermediate between the two. We can understand the differences between these elements in light of \autoref{fig:mass_weighted_metallicity_vs_temperature} as follows. First, note that the sensitivity of the spectrum to oxygen is largely through the prominent O~\textsc{viii} feature at $\approx 0.65$ keV, and the strength of this feature is strongly influenced by the oxygen ionization balance, with the fraction of oxygen in the O~\textsc{viii} reaching a maximum of $\approx 50\%$ at $T \approx 2.5\times 10^6$ K, and falling below 10\% for $T\lesssim 1.2\times 10^6$ K or $T \gtrsim 6\times 10^6$ K \citep{Sutherland93a}. Consequently, oxygen strongly reflects the abundances prevalent at the lower-temperature end of ``hot gas'', $T \approx (2-3)\times10^{6}$ K. By contrast, the greatest sensitivity to Ne is to Ne~\textsc{ix} and Ne~\textsc{x} features at $\approx 1$ keV, and the latter ionization state accounts for more than 10\% of the available neon only at temperatures $T \approx (2.5 - 10)\times 10^6$ K, with a maximum of $\approx 50\%$ abundance at $T\approx 5\times 10^6$ K \citep{Sutherland93a}. This makes Ne sensitive to significantly hotter gas than O, and so when abundant very hot gas is present -- as \autoref{fig:mass_weighted_metallicity_vs_temperature} shows it is in regions A1 and A3 -- the metallicity inferred from Ne winds up reflecting the higher metallicity of this gas. By contrast iron is somewhat more complex, since the X-ray spectral feature most sensitive to iron is the complex of Fe L lines at $\approx 1$ keV, which can be produced by Fe in ionization states from Fe~\textsc{xviii} to Fe~\textsc{xxiv}. The lowest of these ionization states reaches $\gtrsim 10\%$ abundance at $T \approx 3 \times 10^6$ K, leading to a behaviour that is somewhat intermediate between that of O and Ne, although the temperature dependence is more complex for Fe because the highest ionization state that produces Fe L emission remains abundant up to much higher temperatures, $T\approx 3\times 10^7$ K. The net effect of this complex dependence is that in the outer regions (e.g., A3 and A5) the oxygen abundance is closer to the X-ray weighted metallicity $Z_\mathrm{met}$, while near the disc (e.g., A1) the iron abundance is closer.

In summary, we can understand qualitatively why different elements are biased by different amounts based on the fact that they are sensitive to different ranges of gas temperature, and that gas at different temperatures really does have systematically different metallicities due to the effects of cooling by dilution. Thus, while metallicity declines with distance from the midplane, and all elements properly capture this trend, the discrepancy between the abundances derived for any particular element and the ``true'' metallicity value arises naturally from these complex combinations of atomic physics and the presence of a metallicity-temperature correlation in the gas. Given the complexity of this dependence, it seems unlikely that we will be able to acquire significantly more accurate results from spatially-unresolved spectra.

\section{Conclusion}
\label{sec:conclusion}

We study mock X-ray emission from galactic winds generated in a high-resolution hydrodynamic simulation of the Solar Neighbourhood taken from the QED simulation suite, with particular attention to spatial variations in the elemental abundances across the wind derived from analysis of the X-ray spectra. The simulations possess a significant gradient in abundances whereby the hot gas closer to the galactic plane, which is dominated by fresh supernova ejecta, is more metal-rich than the hot gas further from the plane, which has been diluted down by mixing with more metal-poor cool gas entrained into the wind. Our analysis reveals that this effect should lead to a metallicity gradient that is detectable from X-ray spectra, with higher abundances of $\alpha$ elements like oxygen and silicon observed closer to the disc compared to the outer regions of the outflows. Such a pattern has already been observed in the wind of M82, and our simulations provide a natural explanation for it.

Our main conclusions are:
\begin{enumerate}
    \item Gradients in elemental abundance inferred from X-ray spectra of galactic winds, with a peak at the center and decline at high altitudes, are a result of incomplete mixing between the hot and cold phases.
    \item Conversely, the presence of such gradients in observations is strong evidence for preferential metal loading of winds -- in the language of galactic chemical evolution models, they are evidence that the yield reduction factor $\phi$, which measures the fraction of SN ejecta that are retained in the disc and available to enrich the next generation of stars, is $< 1$. Models that are not metal-loaded do not reproduce observations.
    \item The strength of the gradient can be used as a diagnostic for the strength of evaporation of cold gas entrained by winds into the hot phase, with a steeper gradient implying an increasing contribution to the mass of the hot gas phase by evaporation of cool clouds.
    \item Making these measurements in practice is quite sensitive to the signal-to-noise ratio (SNR) of the X-ray data; low SNR data can create artificial gradients or mask real ones. We find an $\textsc{SNR}\gtrsim30$ is necessary to faithfully extract metallicity values from the spectrum, and that $\textsc{SNR}\gtrsim 10$ is sufficient to detect the presence of a gradient, but probably not to recover its shape faithfully.
    \item Even with high SNR data, however, there are likely to be non-negligible systematic errors in inferred gradients arising from the need to assume a single set of abundances for all of the gas in order to avoid the number of free parameters in the fit growing too large. In the most poorly-mixed regions close to the disc, which contain bubbles of fresh SN ejecta with very high metallicity and temperature, this assumption breaks down.
\end{enumerate}


\section*{Software}

This research made use of \texttt{numpy} \citep[\url{https://numpy.org}]{numpy}, \texttt{matplotlib} \citep[\url{https://matplotlib.org/}]{matplotlib}, \texttt{yt} \citep[\url{https://yt-project.org/}]{turk11}, \texttt{pyXSIM} \citep[\url{http://www.ascl.net/1608.002}]{zuhone16}, \texttt{SOXS} \citep[\url{http://ascl.net/2301.024}]{zuhone23} and \textsc{Quokka} \citep[\url{https://github.com/quokka-astro/quokka}]{wibking22, He24a}.

\section*{Acknowledgements}

AV and MRK acknowledge support from the Australian Research Council through awards FL220100020 and DP230101055. The simulations suite QED and the mock data discussed in this paper were generated with the assistance of resources from the National Computational Infrastructure (NCI Australia), an NCRIS enabled capability supported by the Australian Government, and from the Pawsey Supercomputing Research Centre's Setonix Supercomputer (\url{https://doi.org/10.48569/18sb-8s43)}, with funding from the Australian Government and the Government of Western Australia. RH thanks Andrew J. Battisti, Tianmu (Tim) Gao, Chong-Chong He, Yunhe Li and Yusen Li for commenting on this work.

\section*{Data Availability}

The software pipeline used in this analysis is available from \url{https://github.com/Rongjun-ANU/Xpipeline}. Due to their large size, the raw QED simulation outputs on which the analysis is performed are not available in this repository, but are available on reasonable request to the authors.




\bibliographystyle{mnras}
\bibliography{astro} 




\begin{appendices}

\appendix

\section{Convergence in Time and Resolution}
\label{app:convergence}

\begin{figure}
    \centering
    \includegraphics[width=\columnwidth]{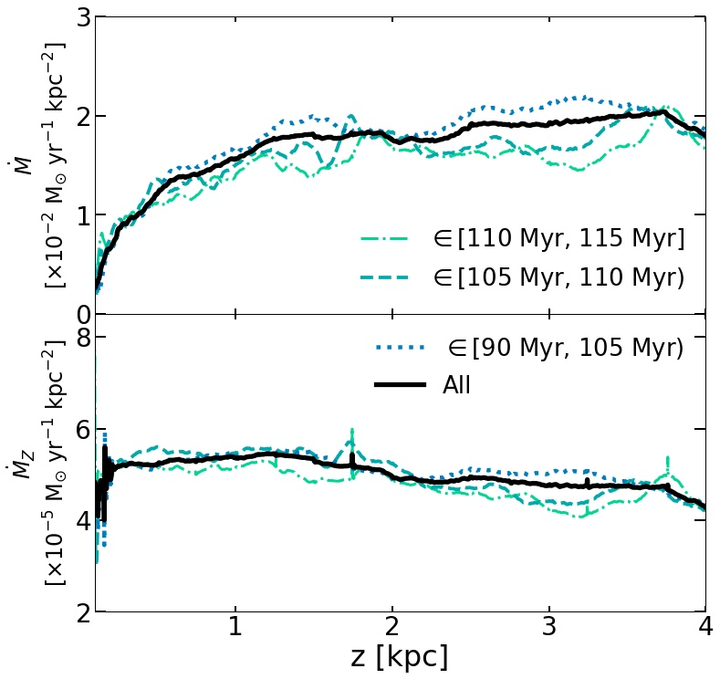}
    \caption{Time-averaged mass outflow rate (upper panel) and metal outflow rate (lower panel) versus height, for several time intervals between 90 and 115 Myr as indicated in the legend.}
    \label{fig:mass_metal_profile_appendix}
\end{figure}

Here we reproduce convergence results from \citetalias{vijayan24} to demonstrate that our simulations have settled to steady-state in time and with respect to the simulation resolution. To evaluate the former, in \autoref{fig:mass_metal_profile_appendix} we plot the mass and metal outflow rates as a function of the distance from the midplane (averaging the $+z$ and $-z$ directions) averaged over three time intervals: $90-105$ Myr, $105-110$ Myr, and $110-115$ Myr. The figure shows that, while there are stochastic fluctuations, as expected given that supernovae are injected stochastically, the outflow rates are fluctuating above the mean and are not systematically increasing or decreasing with time. This indicates that we have reached statistical steady-state.

\begin{figure}
    \centering
    \includegraphics[width=\columnwidth]{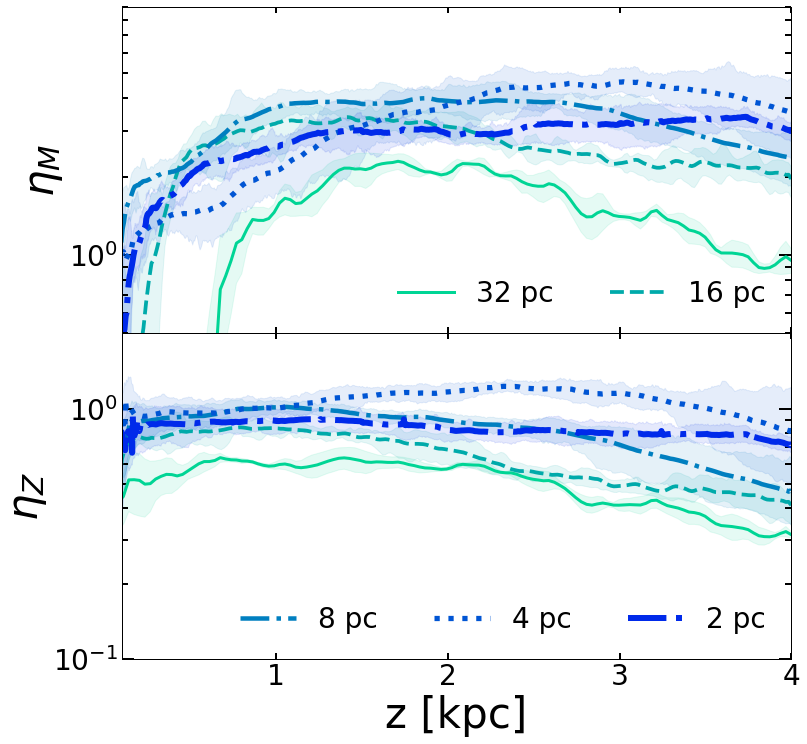}
    \caption{Mass (top panel) and metal (bottom panel) loading factors versus height for different spatial resolutions, measured at times and using averaging intervals identical to those in \citetalias{vijayan24}. Lines represent mean profiles, while shaded bands show the 16th--84th percentile spread over time.}
    \label{fig:convergence_test}
\end{figure}

To check for convergence with respect to resolution, in \autoref{fig:convergence_test} we show profiles of the mass and metal loading factors (defined by normalising the outflow rates to the injection rates -- see \cite{vijayan_prep}) as a function of height for runs at resolutions of 32 pc, 16 pc, 8 pc, 4 pc, and 2 pc, all averaged over the same time interval used in \citetalias{vijayan24} and again averaging the $+z$ and $-z$ directions together. In this plot, lines indicate the median and shaded regions mark the 16th to 84th percentile variation over time. We see that, although there are significant fluctuations (again as expected since supernova events occur stochastically), the direction of change is no longer \emph{monotonic} once the grid spacing reaches $\sim2$ pc. Instead, the profiles oscillate around a well-defined mean, indicating that key outflow properties have converged to within modest fluctuations.

\end{appendices}


\bsp	
\label{lastpage}
\end{document}